\documentclass[12pt,amsmath,amssymb,nofootinbib,superscriptaddress]{revtex4-1}
\usepackage{amsmath} %
\usepackage{pstricks}
\usepackage{color} 
\usepackage{amssymb}
\usepackage{slashed}
\usepackage{graphicx,amsfonts,layout,appendix,subfigure}
\input{epsf} 
\usepackage{breqn}
\allowdisplaybreaks

\usepackage{verbatim} 
\usepackage{upgreek}
\usepackage{microtype}
\def\gev{\,\text{GeV}^2}
\def\beq{\begin{equation}}
\def\eeq{\end{equation}}
\def\bea{\begin{eqnarray}}
\def\eea{\end{eqnarray}}
\def\beqn{\begin{eqnarray}} 
\def\eeqn{\end{eqnarray}}

\def\nn{\nonumber}

\def\ln#1{\mathrm{ln}\left(#1\right)}
\def\lnn#1{\mathrm{ln^2}\left(#1\right)}
\def\lnnn#1{\mathrm{ln^3}\left(#1\right)}
\def\li#1{\mathrm{Li_2}\left(#1\right)}

\newcommand\alphas{\alpha_{\mathrm{S}}}
\newcommand\as{a_{\mathrm{S}}}
\newcommand\al{a}

\newcommand\aso{a_{\mathrm{S}, 0}}
\newcommand\alo{a_0}

\newcommand{\m}[1]{\mathbf{#1}}

\def\beq{\begin{equation}} \def\eeq{\end{equation}}
\def\beqn{\begin{eqnarray}} \def\eeqn{\end{eqnarray}}
 
\def\nn{\nonumber}

\begin{document} 

\begin{titlepage}
\renewcommand{\thefootnote}{\fnsymbol{footnote}}
\begin{flushright}
     \end{flushright}
\par \vspace{10mm}

\begin{center}
{\Large \bf
Analytical solution for QCD $\otimes$ QED evolution}
\end{center}

\par \vspace{2mm}
\begin{center}
{\bf Daniel de Florian}~$^{(a)}$\footnote{{\tt deflo@unsam.edu.ar}}, {\bf Lucas Palma Conte}~$^{(a)}$\footnote{{\tt lpalmaconte@unsam.edu.ar}}

\vspace{5mm}

${}^{(a)}$
International Center for Advanced Studies (ICAS), ICIFI \& ECyT-UNSAM, 25 de Mayo y Francia,
(1650) San Mart\'\i n,  Pcia. Buenos Aires, Argentina

\vspace{5mm}

\end{center}

\par \vspace{2mm}
\begin{center} {\large \bf Abstract} \end{center}
\begin{quote}
\pretolerance 10000


We present an analytical solution for the evolution of parton distributions incorporating mixed-order QCD $\otimes$ QED corrections, addressing both polarized and unpolarized cases. Using the Altarelli-Parisi kernels extended to mixed order, we solve the DGLAP equations exactly in Mellin $N$-space and derive the associated Wilson coefficients for the polarized structure function $g_1$. Our analytical approach improves computational efficiency and enhances the precision of theoretical predictions in observables sensitive to QED corrections, with relevance for current and future phenomenological applications.


\end{quote}
\begin{flushleft}

\end{flushleft}
\end{titlepage}

\setcounter{footnote}{0}


\setcounter{footnote}{0}
\renewcommand{\thefootnote}{\fnsymbol{footnote}}

\section{INTRODUCTION}
\label{sec:introduction}
Increasing the precision of theoretical calculations for partonic cross-sections is essential. QCD computations have reached remarkable levels of accuracy, even up to next-to-next-to-next-to-leading order (N$^3$LO) in some cases. As a result, additional effects that were previously neglected, such as QED corrections, are beginning to play a more significant phenomenological role in theoretical predictions. Achieving a comprehensive description at this level for any hadronic process requires matching accuracy on the non-perturbative side as well. In particular, it is crucial to precisely determine Parton Distribution Functions (PDFs) and, in the polarized case, polarized Parton Distribution Functions (pPDFs).

Since PDFs and pPDFs cannot be derived from first principles, they must be extracted from experimental data through global fits. However, their energy evolution can be computed using the Dokshitzer–Gribov–Lipatov–Altarelli–Parisi (DGLAP) equations. The solution of these equations relies on knowledge of the Altarelli–Parisi splitting functions (AP kernels) \cite{Altarelli:1977zs}.

On the unpolarized side, the computation of NNLO corrections to the splitting functions, as presented in \cite{Moch:2001im, Moch:2004pa, Vogt:2004mw, Vogt:2005dw}, along with the development of modern parton distribution analyses \cite{NNPDF:2021njg, Hou:2019efy, Bailey:2020ooq, H1:2021xxi, PDF4LHCWorkingGroup:2022cjn}, have enabled the level of accuracy required for QCD phenomenology. 

Furthermore, the LO QED corrections for the kernels were computed in \cite{Roth:2004ti, Kripfganz:1988bd,Spiesberger:1994dm}. A significant advancement in this field was presented in \cite{deFlorian:2016gvk, deFlorian:2015ujt}, where the so-called ``Abelianization" algorithm was introduced. This algorithm allows the computation of NLO QED and mixed-order $\mathcal{O}(\alphas\alpha)$ corrections based on the known NNLO and NLO results in QCD, respectively. Solutions to the DGLAP evolution equations incorporating LO QED corrections were provided in \cite{Roth:2004ti, Bertone:2013vaa, Sadykov:2014aua}, while several global fits of parton distributions including these corrections were performed in \cite{Carrazza:2015dea, Martin:2004dh, Ball:2013hta}.
When QED corrections are incorporated into PDF analyses, the photon PDF becomes a crucial component. Several studies have attempted to extract this density from data, but most rely on phenomenological models, introducing significant uncertainties in the fit \cite{Martin:2004dh, Ball:2013hta, Schmidt:2015zda, Harland-Lang:2016kog}. In this context, the LuxQED framework \cite{Manohar:2016nzj, Manohar:2017eqh} offers a novel method for computing the photon PDF in terms of the proton's structure functions, enhancing both accuracy and reducing reliance on phenomenological assumptions.

On the other hand, for the polarized case, the polarized QCD splitting functions are known at LO \cite{Altarelli:1977zs}, NLO \cite{Mertig:1995ny,Vogelsang:1995vh,Vogelsang:1996im} and NNLO \cite{Moch:2014sna,Vogt:2014pha,Blumlein:2021enk}. Several global analyses were performed to fit the pPDFs at NLO, including DSSV \cite{deFlorian:2009vb,deFlorian:2014yva, DeFlorian:2019xxt}, NNPDF \cite{Nocera:2014gqa}, JAM collaboration \cite{Ethier:2017zbq}, and recently at NNLO \cite{Bertone:2024taw,Borsa:2024mss,Cruz-Martinez:2025ahf}.
To incorporate QED corrections, the LO QED contributions to the polarized splitting functions, along with the analytical solution of the DGLAP equations including these corrections, were introduced in \cite{deFlorian:2023zkc}. Also, a calculation of the polarized photon distribution function based on the LuxQED approach was presented in \cite{deFlorian:2024hsu}. In \cite{Rein:2024fns}, it was shown that the photon PDF - and consequently QED corrections - play a significant role in the computation of single-inclusive prompt photon production in electron–proton collisions.

In this study, our goal is to analytically solve the DGLAP equations at $\mathcal{O}(\alphas\alpha)$ for both the unpolarized and polarized cases. The analytical approach simplifies the implementation and results in faster computations, which is especially useful for repeated evaluations in phenomenological studies. For the polarized case, we present the mixed-order corrections to the polarized AP kernels, computed by applying the ``Abelianization" algorithm to the NLO kernels. To solve the DGLAP equations, we propose two methods. The first is an extension of the conventional $U$-matrix approach \cite{Vogt:2004ns}, incorporating QCD $\otimes$ QED corrections. The second method builds on a development \cite{Simonelli:2024vyh} that introduced a novel strategy to solve the NLO QCD equations using the Magnus expansion \cite{magnus1954, BLANES2009151}, which has also been applied in a recent work to obtain semi-analytical solutions in $x$-space \cite{Haug:2024asl}. With this second method, we explore the applicability of the Magnus approach to the QCD $\otimes$ QED evolution.

Lastly, we compute the mixed-order corrections to the structure function $g_1$, derived from the known NNLO QCD Wilson coefficients \cite{Zijlstra:1993sh} using the ``Abelianization" technique, and study their phenomenological consequences. 

The paper is organized as follows. In Section \ref{sec:basis} we present the evolution equations and
establish the main notation for this article as well as the basis on which we will work. In Section \ref{sec:splitting-kernels}, we introduce the polarized splitting kernels at mixed-order. In Section \ref{sec:mixsolu}, we first present a solution for the Renormalization Group Equations (RGE) at mixed order and discuss some features of Mellin space. Then, we solve the DGLAP equations in both the singlet and non-singlet cases. The polarized structure function $g_1$ with mixed-order corrections is presented in Section \ref{sec:g1}. Finally, in Section \ref{sec:conclusions}, we present the conclusions of this work.

\section{Splitting kernels and parton distribution basis}
\label{sec:basis}

We begin by writing the general expressions for the evolution of gluon, photon, and quark distributions. We will write all equations for the unpolarized case. The polarized case is obtained by simply considering the corresponding polarized proton distribution
and splitting functions, i.e. basically adding $\Delta$ in all the expressions below.
\beqn
\frac{dg}{dt}&=& \sum_{j=1}^{n_f} P_{g q_j} \otimes  q_j+ \sum_{j=1}^{n_f} P_{g \bar{q}_j} \otimes \bar{q}_j +  P_{g g} \otimes g +  P_{g \gamma} \otimes \gamma \, , \nn
\\ \frac{d\gamma}{dt}&=& \sum_{j=1}^{n_f} P_{\gamma q_j} \otimes  q_j+ \sum_{j=1}^{n_f} P_{\gamma \bar{q}_j} \otimes \bar{q}_j + P_{\gamma g} \otimes g + P_{\gamma \gamma} \otimes \gamma \, ,\label{Eq:DGLAP}
\\\frac{dq_i}{dt}&=& \sum_{j=1}^{n_f} P_{q_i q_j} \otimes q_j + \sum_{j=1}^{n_f} P_{q_i \bar{q}_j} \otimes \bar{q}_j + P_{q_i g} \otimes g + P_{q_i \gamma} \otimes \gamma \, ,\nn
\eeqn
with $t=\ln{Q^2}$ ($Q$ being the factorization scale), $n_f$ is the number of active fermions and $P_{ij}$ the Altarelli-Parisi splitting functions in the space-like region. The analogue evolution equations for antiquarks can be obtained by applying charge conjugation. Here, we use the conventional notation
\beq
(f\otimes g)(x) = \int_x^1 \, \frac{dy}{y} \, f\left(\frac{x}{y}\right)g(y) \, ,
\eeq
to indicate convolutions. We do not include the lepton distributions, since up to the order we reach here they basically factorize from the rest of the distributions\footnote{To $\mathcal{O}(\alpha)$ lepton distributions only couple, in a trivial way, to the photon density.}.
Along this work we will use the expressions for the splitting functions including QCD, QED and mixed-order perturbative corrections. In this sense, each kernel can be expressed as,
\beqn
P_{ij} &=&\sum_{\{n,m\}} \as^n \, a^m \, P_{ij}^{(n,m)},
\label{eq:expanP}
\eeqn
where the indices $(n,m)$ indicate the (QCD,QED) perturbative order of the calculation, with $\as \equiv \frac{\alphas}{4\pi}$ the strong coupling and $a \equiv \frac{\alpha}{4\pi}$ the electroweak coupling. Due to the QED corrections, the kernels can depend on the electric charge of the initiating quarks (up or down type). In general, we have $P_{i,j}^{(n,1)} \sim e_q^2$ with at least some $i,j=q$.

The quark-splitting functions are decomposed as
\beqn
P_{q_i \, q_k} &=& \delta_{ik} \, P^V_{qq} + P^S_{qq} \, ,\\
 P_{q_i \, \bar{q}_k} &=& \delta_{ik} \, P^V_{q\bar{q}} + P^S_{q\bar{q}} \, ,\\
  P_q^{\pm} &=& P^V_{qq} \pm P^V_{q\bar{q}} \, ,
  \label{Eq:Pv}
\eeqn
which act as a definition for $P^V_{q q}$ and $P^V_{q \bar{q}}$. To minimize the mixing between the different parton distributions in the evolution, it is convenient to introduce the following basis, which distinguishes the singlet and the non-singlet PDF combinations \cite{Roth:2004ti}:
\beqn
\label{eq:basis}
f^{NS}&=&\{u_v,d_v,s_v,c_v,b_v,
\Delta_{uc},\Delta_{ds},\Delta_{sb},\Delta_{ct}\}, \nn \\
 f^{S}&=&\{\Delta_{UD}, \Sigma,g,\gamma\ \} ,
\eeqn
where
\beqn
q_{v_i} &=& q_i-\bar{q_i} \, , \\
\Delta_{ct} &=& c+\bar{c}-t-\bar{t} \, ,\nn \\ 
\Delta_{uc} &=& u+\bar{u}-c-\bar{c} \, ,\nn \\ 
\Delta_{ds} &=& d+\bar{d}-s-\bar{s} \, ,\nn \\ 
\Delta_{sb} &=& s+\bar{s}-b-\bar{b} \, , \\ 
\Delta_{UD} &=& u+\bar{u}+c+\bar{c} -d-\bar{d} -s-\bar{s}-b-\bar{b} \, , \\
 \Sigma &=& \sum_{i=1}^{n_f} ( q_i+\bar{q}_i)  \, .
\eeqn
$\Delta_{UD} $ could also include the top quark distribution in the case of a 6 flavour analysis (adding $\Delta_{ct}$ and $t_v$ to complete the basis). 

Taking into account that beyond NLO in QCD the singlet {\it non-diagonal terms} ($P^S_{q \bar{q}}$ and $P^S_{qq}$) differ \cite{Catani:2004nc}, it is useful to define 
\beqn
\Delta P^S & \equiv & P^S_{qq} -P^S_{q \bar{q}}  , \nn \\
 P^S & \equiv &  P^S_{qq} + P^S_{q \bar{q}}, 
 \label{Eq:Ps}
\eeqn
where we explicitly use that these contributions do not depend on the quark charge up to the order we reach, since they do not receive QED corrections to $\mathcal{O}(\al)$.
In the expansion in powers of the couplings, the flavour-diagonal (‘valence’) quantity $P_{qq}^V$ in Eq.(\ref{Eq:Pv}) begins at first order. On the other hand, $P_{q\bar{q}}^V$ and the flavour-independent (‘sea’) contributions $P^S_{qq}$ and $P^S_{q\bar{q}}$, and therefore the ‘pure-singlet’ term $P^S$ in Eq.(\ref{Eq:Ps}), are of order $\alpha_s^2$. The kernel $\Delta P^S $ in Eq.(\ref{Eq:Ps}) arises for the first time at the third order in QCD.
The evolution equations Eq.(\ref{Eq:DGLAP}) for the parton distributions in the basis of Eq.(\ref{eq:basis}) read for the non-singlet sector,
\beqn
\frac{dq_{v_i}}{dt} &=&   P_{q_i}^-     \otimes q_{v_i}  +\sum_{j=1}^{n_f} \Delta P^S  \otimes  q_{v_j}    \, ,
\label{eq:evolucionqvSIMPLE}
\\ \frac{d \{ \Delta_{uc} , \Delta_{ct} \}}{dt} &=&  P_{u}^+ \otimes \{ \Delta_{uc} , \Delta_{ct}  \} \, ,
\label{eq:evolucionDupperSIMPLE}
\\ \frac{d \{ \Delta_{ds} ,\Delta_{sb}  \}}{dt} &=&  P_{d}^+ \otimes \{ \Delta_{ds} ,\Delta_{sb}  \} \, ,
\label{eq:evolucionDlowerSIMPLE}
\eeqn
where in the second and third lines we use the notation $P_{u(d)}$, differentiating the kernels in two groups, those corresponding to the {\it up} quark flavours  ($u, \, c, \, t $)  and those corresponding to the {\it down} quark flavours ($d, \, s, \, b$).
As noted above, the term $\Delta P^S$ is zero up to the order we work; therefore, in the non-singlet sector we have decoupled equations for the distributions $q_{v_i}$, i.e., the last term in Eq.(\ref{eq:evolucionqvSIMPLE}) vanish. For the singlet sector, we have, 
\beqn
\frac{d  \Delta_{UD}  }{dt} &=&  \frac{P_{u}^+ + P_{d}^+}{2} \otimes \Delta_{UD} + \frac{P_{u}^+ - P_{d}^+}{2} \otimes \Sigma +  (n_u-n_d) P^S \, \otimes \Sigma\nn 
\\ &+& 2 (n_u P_{ug} -n_d P_{dg})  \otimes g + 2 (n_u P_{u\gamma} -n_d P_{d\gamma})  \otimes \gamma \, ,\label{eq:evolucionDUDSIMPLE}
\\ \nn 
\frac{d  \Sigma  }{dt} &=&  \frac{P_{u}^+ + P_{d}^+}{2} \otimes \Sigma+ \frac{P_{u}^+ - P_{d}^+}{2} \otimes \Delta_{UD} +  n_f \, P^S \otimes \Sigma\nn 
\\ &+& 2 (n_u P_{ug} +n_d P_{dg})  \otimes g + 2 (n_u P_{u\gamma} +n_d P_{d\gamma})  \otimes \gamma \, ,\label{eq:evolucionSIGMASIMPLE}
\\
\frac{d  g  }{dt} &=& \frac{P_{g u}-P_{g d}}{2} \otimes \Delta_{UD} +\frac{P_{g u}+P_{g d}}{2}\otimes \Sigma +P_{g g} \otimes g +P_{g \gamma} \otimes \gamma \,,\label{eq:evolucionGLUONSIMPLE}
\\
\frac{d  \gamma  }{dt} &=&\frac{P_{\gamma u}-P_{\gamma d}}{2} \otimes \Delta_{UD} +\frac{P_{\gamma u}+P_{\gamma d}}{2}\otimes \Sigma+P_{ \gamma g} \otimes g + P_{\gamma\gamma} \otimes \gamma  \, .\label{eq:evolucionGAMMASIMPLE}
\eeqn
Notice that in the limit of an equal number of $u$ and $d$ quarks ($n_u = n_d$ ) and the same electric charges ($P_{ug} = P_{dg}$ , $P_{u\gamma} = P_{d\gamma}$ , $P_u^+ = P_d^+$ ), $\Delta_{UD}$  decouples from the other distributions in the evolution, while the singlet evolution recovers the usual pure QCD expression.
\section{Explicit Formulae for the Splitting Kernels}
\label{sec:splitting-kernels}
To solve the DGLAP equations up to $\mathcal{O}(\as\, \al)$, it is necessary to employ AP kernels computed at the same level of accuracy. In \cite{deFlorian:2015ujt}, a method known as ``Abelianization" was introduced to compute mixed-order corrections of the AP kernels from NLO in QCD. In the same work, the unpolarized kernels were computed. For the polarized kernels, we applied the same method to the results in \cite{Mertig:1995ny,Vogelsang:1995vh}. All results are obtained in the $\overline{\text{MS}}$ Modified Minimal Subtraction scheme, and the treatment of $\gamma^5$ is performed using the HVBM scheme. In the first place, we obtain the photon-initiated processes,
\beqn
\Delta P_{q\gamma}^{(1,1)} &=& 2\,C_F \, C_A \, e_q^2 \Bigg\{ -22+27 x -9 \ln{x} +8(1-x) \ln{1-x} \nn\\
&+&  \delta p_{qg}(x) \left[  2\lnn{1-x}- 4\ln{1-x} \ln{x}+\lnn{x} -\frac{2\pi^2}{3} \right]\Bigg\}  \, ,\label{eq:kernels1}\\  
\Delta P_{g\gamma}^{(1,1)} &=& -4\,C_F \, C_A \, \left(\sum_{j=1}^{n_F} e_{q_j}^2\right)\,\Bigg\{10(1-x) +2(5-x)\ln{x}+2(1+x)\lnn{x} \Bigg\}, \label{eq:kernels2}\\
\Delta P_{\gamma\gamma}^{(1,1)} &=& -4\,C_F \, C_A \, \left(\sum_{j=1}^{n_F} e_{q_j}^2\right)  \delta(1-x), \label{eq:kernels3}
\eeqn
where $\delta p_{qg}(x)=4x-2$. For processes with a starting gluon, we obtain,
\beqn
\Delta P_{qg}^{(1,1)} &=&  2\,T_R \, e_q^2 \Bigg\{ -22+27 x -9 \ln{x} +8(1-x)\ln{1-x}\nn \\
&+& \delta p_{qg}(x) \left[  2\lnn{1-x}- 4\ln{1-x} \ln{x}+\lnn{x} -\frac{2\pi^2}{3}\right]\Bigg\}  \, ,\label{eq:kernels4}\\
\Delta P_{\gamma g}^{(1,1)} &=& 4\,T_R \,\left(\sum_{j=1}^{n_F} e_{q_j}^2\right) \,\Bigg\{10(1-x) +2(5-x)\ln{x}+2(1+x)\lnn{x} \Bigg\}\, , \label{eq:kernels5}\\
\Delta P_{gg}^{(1,1)} &=& -4\,T_R \,\left(\sum_{j=1}^{n_F} e_{q_j}^2\right) \, \delta(1-x) \, .\label{eq:kernels6}
\eeqn
Notice that the corrections to the splitting kernels $\Delta P_{\gamma \gamma}$ and $\Delta P_{gg}$ are proportional to the Dirac`s delta function $\delta (1-x)$ since they are originated by virtual two-loop contributions to the photon and gluon propagators, respectively. The splitting functions with an initial quark are given by 
\beqn
\Delta  P_{gq}^{(1,1)} &=&4\, e_q^2\,C_F \,  \Bigg\{ (-4 -\lnn{1-x}+\frac{1}{2}\lnn{x})\delta p_{gq}(x) -\frac{1}{2}- \frac{1}{2}(4-x)\ln{x}\nn\\&-& \delta p_{gq}(-x) \ln{1-x}  \Bigg\}  ,\label{eq:kernels7}\\
\Delta P_{\gamma q}^{(1,1)}&=&\Delta P_{g q}^{(1,1)}.\label{eq:kernels8}
\eeqn
where $\delta p_{gq}(x) \equiv 2 - x $.
Lastly, due to the conservation of the helicity in a quark line, the quark splitting functions $\Delta P_{qq}^{S(1,1)}$, $\Delta P_{q\bar{q}}^{S(1,1)}$,$\Delta P_{qq}^{V(1,1)}$ and $\Delta P_{q\bar{q}}^{V(1,1)}$ coincide with those in the unpolarized case computed in \cite{deFlorian:2016gvk, deFlorian:2015ujt},
\beqn
\Delta P_{qq}^{S(1,1)} &=& \Delta P_{q\bar{q}}^{S(1,1)}=0 \, , 
\\ \nn \Delta P_{qq}^{V(1,1)} &=& -  8 \, C_F \, e_q^2 \left[\left(2 \ln{1-x}+\frac{3}{2}\right)\ln{x} p_{qq}(x) + \frac{3+7x}{2}\ln{x} + \frac{1+x}{2}{\lnn{x}}  \right.
\\ &+& \left. 5(1-x) + \left( \frac{\pi^2}{2}-\frac{3}{8}-6 \zeta_3 \right) \delta(1-x)  \right] \, , \label{eq:kernels10}
\\ \Delta P_{q\bar{q}}^{V(1,1)} &=&  8\, C_F \, e_q^2 \left[4(1-x)+2(1+x)\ln{x} + 2p_{qq}(-x)S_2(x)\right] \, ,
\label{eq:kernels11}
\eeqn
where $p_{qq}(x)=\frac{1+x^2}{(1-x)_+}$, with the usual plus distribution defined as
$\int_0^1 dx \, \frac{f(x)}{(1-x)_+} = \int_0^1 dx \, \frac{f(x)-f(1)}{1-x} \, ,$
for any regular test function $f$ and, the function $S_2(x)$ is given by,
\beqn
\nn S_2(x) &=& \int_{\frac{x}{1+x}}^{\frac{1}{1+x}} \, \frac{dz}{z} \, \ln{\frac{1-z}{z}} = \li{-\frac{1}{x}} - \li{-x}
\\ &+&  \lnn{\frac{x}{1+x}}-\lnn{\frac{1}{1+x}}\,.
\label{eq:S2definicion}
\eeqn
Singlet contributions, $\Delta P_{qq}^{S(1,1)}$ and $\Delta P_{q\bar{q}}^{S(1,1)}$, are null at this order, as anticipated in Sec. \ref{sec:basis}. 

\section{Mixed solution to DGLAP equations}
\label{sec:mixsolu}
\subsection*{RGE equations at mixed-order}
\begin{figure}
   \centering
    \includegraphics[width=0.7\textwidth]{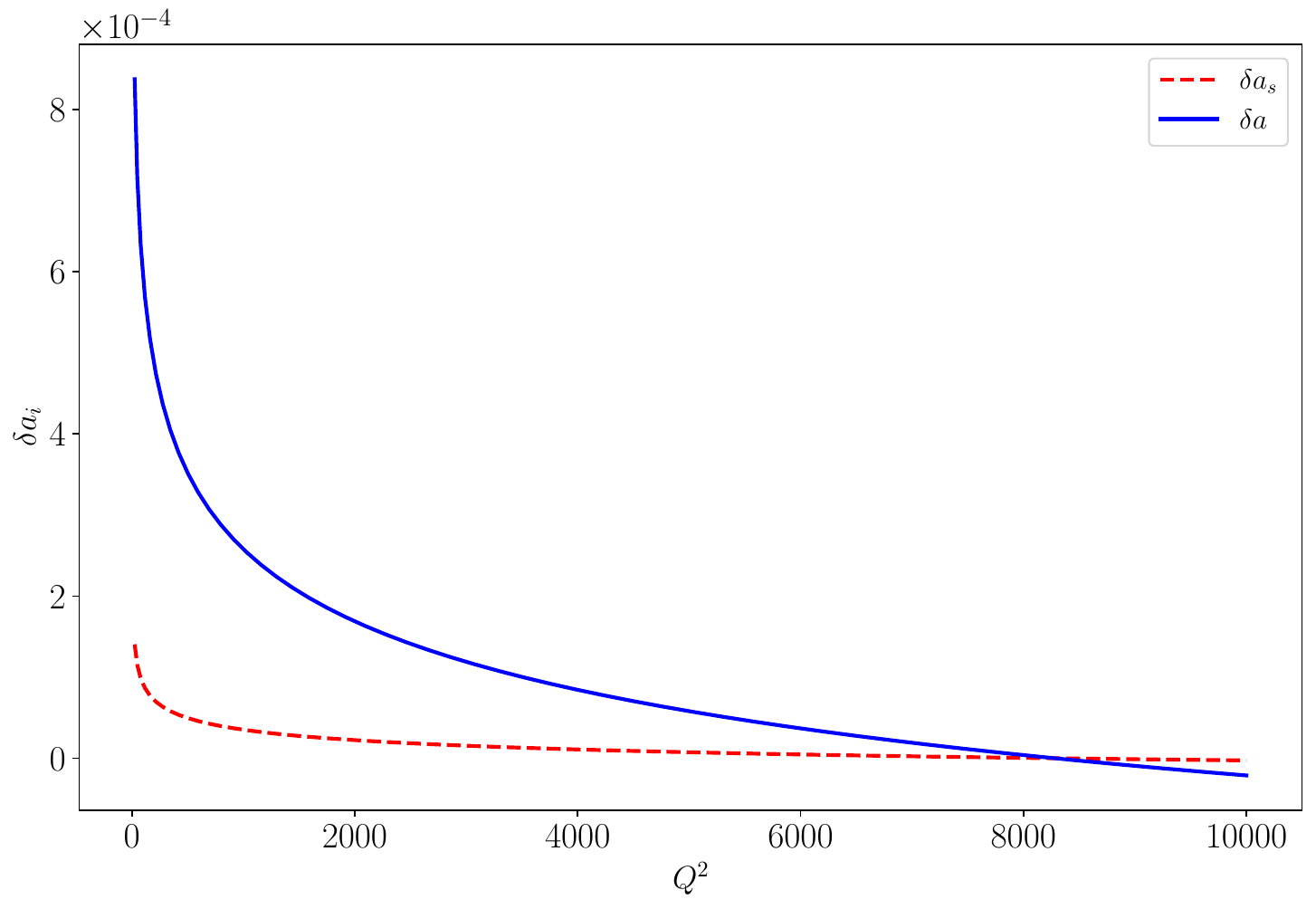}
    \caption{Relative mixed-order corrections to the couplings, as defined in Eq.(\ref{eq:realtivecoupl}). We set the initial scale to $Q^2_0=M^2_z$, where $M_z$ is the mass of the $Z$ boson. For the initial values of the couplings we use $\as(M^2_z)\simeq 0.00946$ \cite{ParticleDataGroup:2014cgo,Deur:2016tte} and $\al(M^2_z)\simeq 0.000610$ \cite{Bouchendira:2010es}
    . We perform the evolution towards lower values of $Q^2$, evolving the couplings backwards accordingly.}
    \label{fig:coupling}
\end{figure}

To solve the DGLAP evolution equations in the mixed case, it is necessary to obtain the evolution of the couplings at this order. The equations governing the energy behaviour of the coupling are the well-known RGE equations. Up to mixed-order they read,
\beqn
\frac{d\as(t)}{dt}&=&-\beta_{10} \as^2(t)-\beta_{20}\as^3(t)-\beta_{11}\as^2(t) \al(t), \nn\\
\frac{d\al(t)}{dt}&=&-\beta'_{01} \al^2(t)-\beta'_{11}\al^2(t) \as(t),
\label{eq:coupling}
\eeqn
where $t=\ln {Q^2}$, and the functions $\beta$ and $\beta'$ are the beta functions for QCD and QED respectively. Notice that the QCD $\otimes$ QED corrections to these functions are required. They can be found in \cite{Cieri:2018sfk}. To solve the system, we can propose a solution of the form,
\beqn
\as&=&U^{\text{QCD}}U^{\text{MIX}}_{1} \aso,\label{eq:couplingevoloperadoresQCD}\\
\al&=&U^{\text{QED}}U^{\text{MIX}}_2 \alo,
\label{eq:couplingevoloperadoresQED}
\eeqn
where $\aso=\as(t_0)$ and $\al_0=\al(t_0)$, with $t_0=\ln{Q^2_0}$. The operators $U^{\text{QCD}}$ and $U^{\text{QED}}$ are the operators that evolve the coupling in QCD and QED, respectively; they can be determined using the RGE equations without mixed terms. They read, 
\beqn
U^{\text{QCD}}&=&\frac{\beta_{10}+\aso \beta_{10}^2 \Delta t-\aso \beta_{20} \ln{1+\aso \beta_{10}\Delta t}}{\beta_{10}(1+\aso \beta_{10}\Delta t)^2}, \label{eq:opevocoupling}\\
U^{\text{QED}}&=&\frac{1}{1+\beta'_{01} \alo \,\Delta t}.\nn
\eeqn
where $\Delta t=t-t_0$. 
We use the NLO QCD solution for the strong coupling and the LO one in QED for the electroweak coupling. Including higher-order corrections in the latter would introduce contributions beyond the accuracy considered in this work for the evolution of both the interactions and the parton densities.
On the other hand, the operators $U^{\text{MIX}}_1$ and $U^{\text{MIX}}_2$ encompass the mixed evolution, we propose they assume the following form, 
\beqn
U^{\text{MIX}}_1=1+A_1\,\aso \alo, \nn\\
U^{\text{MIX}}_2=1+A_2\,\aso \alo, \label{eq:mixsolucoupling}
\eeqn
where $A_1$ and $A_2$ are constants to be determined. Keeping terms up to the order we are working in, we obtain the following system of differential equations for $A_1$ and $A_2$,
\beqn
\left(\begin{array}{c}
A_1\\
A_2
\end{array}\right)&\simeq&\left(\begin{array}{cc}
-\aso\beta_{10} U^{\text{QCD}} & 0 \\
0 & -\alo\beta'_{10} U^{\text{QED}} 
\end{array}\right)\,\left(\begin{array}{c}
A_1\\
A_2
\end{array}\right)-\left(\begin{array}{c}
\beta_{11} U^{\text{QCD}}U^{\text{QED}}\\
\beta'_{11} U^{\text{QCD}}U^{\text{QED}}
\end{array}\right),
\eeqn
where we can use the LO approximation for the operators $U^{\text{QCD}}$ and $U^{\text{QED}}$, i.e., setting $\beta_{20}\rightarrow 0$ in Eq.(\ref{eq:opevocoupling}).
Then the solutions for $A_1$ and $A_2$ are, 
\beqn
A_1&=&-\frac{\beta_{11} \ln {1 + \alo \beta'_{01} \Delta t}}{\alo \beta'_{01} + \alo \aso \beta_{10} \beta'_{01} \Delta t},\nn\\
A_2&=&-\frac{\beta'_{11} \ln {1 + \aso \beta_{10} \Delta t}}{\aso \beta_{10} + \alo \aso \beta_{10} \beta'_{01} \Delta t}.
\eeqn
In the Fig.\ref{fig:coupling} we show the relative corrections as a function of the scale $Q^2$ for both QCD (red) and QED (blue) couplings, defined as,
\beqn
\delta \as &=& \frac{\as^{\textit{NLO}}-\as^{\textit{MIX}}}{\as^{\textit{NLO}}}, \nn\\
\delta \al &=& \frac{\al^{\textit{LO}}-\al^{\textit{MIX}}}{\al^{\textit{LO}}}.
\label{eq:realtivecoupl}
\eeqn
where $\as^{\textit{NLO}}$, $\al^{\textit{LO}}$, $a_i^{\textit{MIX}}$ denote the strong coupling evolved at NLO in QCD, the electroweak coupling evolved at LO in QED, and the couplings evolved with mixed-order corrections, respectively. We evolve the coupling backward starting from the $Z$ boson mass, i.e., $Q^2_0=M_z^2$, and use $\as(M^2_z)\simeq 0.00946$ \cite{ParticleDataGroup:2014cgo,Deur:2016tte} and $\al(M^2_z)\simeq 0.000610$ \cite{Bouchendira:2010es} as initial values.
We can see that the mixed-order effects on the QED coupling are larger than those on the QCD coupling. The corrections are of order $\mathcal{O}(10^{-4})$, consistent with the expected size of mixed-order effects. This confirms that including additional higher-order terms in Eq.(\ref{eq:mixsolucoupling}) would introduce contributions beyond the accuracy considered in this work. Furthermore, we have explicitly verified that QCD corrections beyond NLO in the RGE of Eq.~(\ref{eq:coupling}) induce an effect of order $\mathcal{O}(10^{-6})$ at the level of the QED coupling constant, which is well below the accuracy relevant for our phenomenological study.

\subsection*{Mellin $N$-space}
In order to efficiently solve the DGLAP evolution equations (\ref{eq:evolucionqvSIMPLE}, \ref{eq:evolucionDupperSIMPLE}, \ref{eq:evolucionDlowerSIMPLE}, \ref{eq:evolucionDUDSIMPLE}, \ref{eq:evolucionSIGMASIMPLE}, \ref{eq:evolucionGLUONSIMPLE} and \ref{eq:evolucionGAMMASIMPLE}), we now turn to Mellin $N$-space techniques.  We define the Mellin transformation, from Bjorken $x$-space to
complex  $N$-moment space, as
\beqn
f(N)=\int_0^1  dx \, x^{N-1} \, f(x),
\label{Eq:mellin}
\eeqn
and its inverse reads
\beqn
f(x)=\frac{1}{2\pi i}\int_{\textit{C}_N}dN \, x^{-N} \, f(N),
\label{eq:antitransofr}
\eeqn
here $\textit{C}_N$ denotes a suitable contour in the complex $N$ plane that has an imaginary part ranging from $-\infty$ to $\infty$ and that intersects the real axis to the right of the rightmost pole of $f(N)$. In practice, it is beneficial to choose the contour to be bent at an angle  $< \pi/2$ towards the negative real-$N$ axis \cite{Vogt:2004ns,Gluck:1989ze}.
The integration in Eq.(\ref{eq:antitransofr}) can then be performed numerically very efficiently by choosing the values of $N$ as the supports for a Gaussian integration \cite{Vogt:2004ns}.

The advantage of working in this space is that the convolutions appearing in the evolution equations can be written simply as products,
\beq
(f\otimes g)(N)=f(N) \, g(N),
\eeq
which makes it easier to manipulate the expressions and solve the corresponding equations analytically at a given order. To transform the kernels to the Mellin $N$-space we use the MT-1.0 Mathematica package \cite{Hoschele:2013pvt}. From now on, all kernels and PDFs will be expressed in this space, but we will keep the nomenclature a bit loose to avoid complicating the notation. 
\subsection{Non-singlet case}\label{sec:nonsinglet}
We start with the solution for the non-singlet distributions (Eqs.(\ref{eq:evolucionqvSIMPLE}),(\ref{eq:evolucionDupperSIMPLE}),(\ref{eq:evolucionDlowerSIMPLE})).
If we express the kernels with the form of Eq.(\ref{eq:expanP}) up to the order $\mathcal{O}(\as\, \al)$, we have a generic form for the evolution
\beqn
\frac{df(Q)}{dt}=(\as P^{(1,0)}+\as^2 P^{(2,0)}+\al P^{(0,1)} + \as \al P^{(1,1)} ) f(Q),
\label{eq:evolucion}
\eeqn
where $f$ is any distribution of the non-singlet sector, and $P$ is the kernel corresponding to the evolution of that distribution. Proposing a solution with an evolution operator of the form, $f(Q)=E\left(Q, Q_0\right)f(Q_0)$, we obtain
\beqn
\frac{dE\left(Q, Q_0\right)}{dt}=(\as P^{(1,0)}+\as^2 P^{(2,0)}+\al P^{(0,1)} + \as \al P^{(1,1)} ) E\left(Q, Q_0\right).
\label{eq:evolnonsinglet}
\eeqn
The main idea to solve the equations to mixed order will be to reuse the well-known solutions for pure QCD and QED \cite{deFlorian:2023zkc,Vogt:2004ns}, and to propose a new mixed-order operator that contains the evolution at $\mathcal{O}(\as\al)$, such that the final solution takes the following form,
\begin{equation}
E\left(Q, Q_0\right)=E^{QCD}\left(Q, Q_0\right)E^{QED}\left(Q, Q_0\right)E^{MIX}\left(Q, Q_0\right),
\label{eq:solunonsinglet}
\end{equation}
where $E^{\text{QCD}}$ and $E^{\text{QED}}$ are the evolution operators corresponding to pure QCD and QED evolution, respectively, and can be found in \cite{deFlorian:2023zkc,Vogt:2004ns}. $E^{\text{MIX}}$ is the mixed-order operator that we introduce. Notice that this operator must also incorporate the evolution of the couplings with mixed-order corrections, as we will impose in the following.
To derive the differential equation governing $E^{\text{MIX}}$, we begin by writing the evolution equations for the QCD and QED operators explicitly. For pure QCD, we have
\beqn
\frac{dE^{\text{QCD}}}{d\as}=-\frac{1}{\as \beta_{10}} \left[P^{(1,0)} +\as \left( P^{(2,0)}-\frac{\beta_{20}}{\beta_{10}} P^{(1,0)}\right)\right]E^{\text{QCD}},
\label{eq:QCDderas}
\eeqn
and for pure QED,
\beqn
\frac{dE^{\text{QED}}}{d\al}=-\frac{1}{\al \beta_{01}} P^{(0,1)} E^{\text{QED}}.
\label{eq:QEDderal}
\eeqn
We can write these equations in terms of the derivative with respect to $t$, using the differential equations for the couplings Eq.(\ref{eq:coupling}) (keeping only the terms relevant to the precision we want to achieve),
\beqn
\frac{d E^{\text{QCD}} }{dt}&=& \frac{d E^{\text{QCD}}}{d\as} \frac{d\as}{dt}=  \as  
( P^{(1,0)} + \as P^{(2,0)} + \al \frac{\beta_{11}}{\beta_{10}} P^{(1,0)}) E^{\text{QCD}}  \, ,
\label{eq:QCDderit}
\eeqn
and for QED,
\beqn
\frac{d E^{\text{QED}} }{dt}&=& \frac{d E^{\text{QED}} }{d\al} \frac{d\al}{dt}=  \al P^{(0,1)} 
( 1 +  \as \frac{ \beta'_{11}}{\beta'_{01}}) E^{\text{QED}},
\label{eq:QEDderit}
 \eeqn
where the last terms of both equations contain the effect of mixed-order coupling. Given that Eq.(\ref{eq:evolnonsinglet}) has to be fulfilled, the mixed evolution operator must obey the following equation
 \beqn
\frac{d E^{\text{MIX}} }{dt}=  \as \, \al\,  \left( P^{(11)} -\frac{\beta_{11}}{\beta_{10}} P^{(1,0)} 
-\frac{\beta'_{11}}{\beta'_{01}}  P^{(0,1)} \right) E^{\text{MIX}} \equiv   \as \, \al\,   R^{(11)}   E^{\text{MIX}}.
\label{eq:solumixNS}
 \eeqn
Notice that we arrive at a differential equation expressed in terms of the variable $t$, in contrasts with the cases of the pure QCD and QED operators, where the differential equations are written in terms of $\as$ and $\al$, respectively (Eqs.~(\ref{eq:QCDderas}) and (\ref{eq:QEDderal})). Given that for $E^{\text{MIX}}$ both interactions are simultaneously involved, it is unfeasible to express the equation directly in terms of only one of the two couplings, therefore, a change of variables $\as, \al \rightarrow t$ is required\footnote{In \cite{Mottaghizadeh:2017vef} to keep solving everything in terms of a single coupling $\as$, they parametrize $\al$ as a linear function of $\as$, which might work numerically, but is totally inconsistent with the well-known behaviour of the running couplings.}.
The solution to Eq.(\ref{eq:solumixNS}) is,
\beqn
\ln{E^{\text{MIX}}}(Q,Q_0) &=& R^{(1,1)} \int_{t_0}^t \as(t)\, \al(t) \, dt.  \label{eq:Eintegral}
\eeqn

\begin{figure}
   \centering
    \includegraphics[width=1\textwidth]{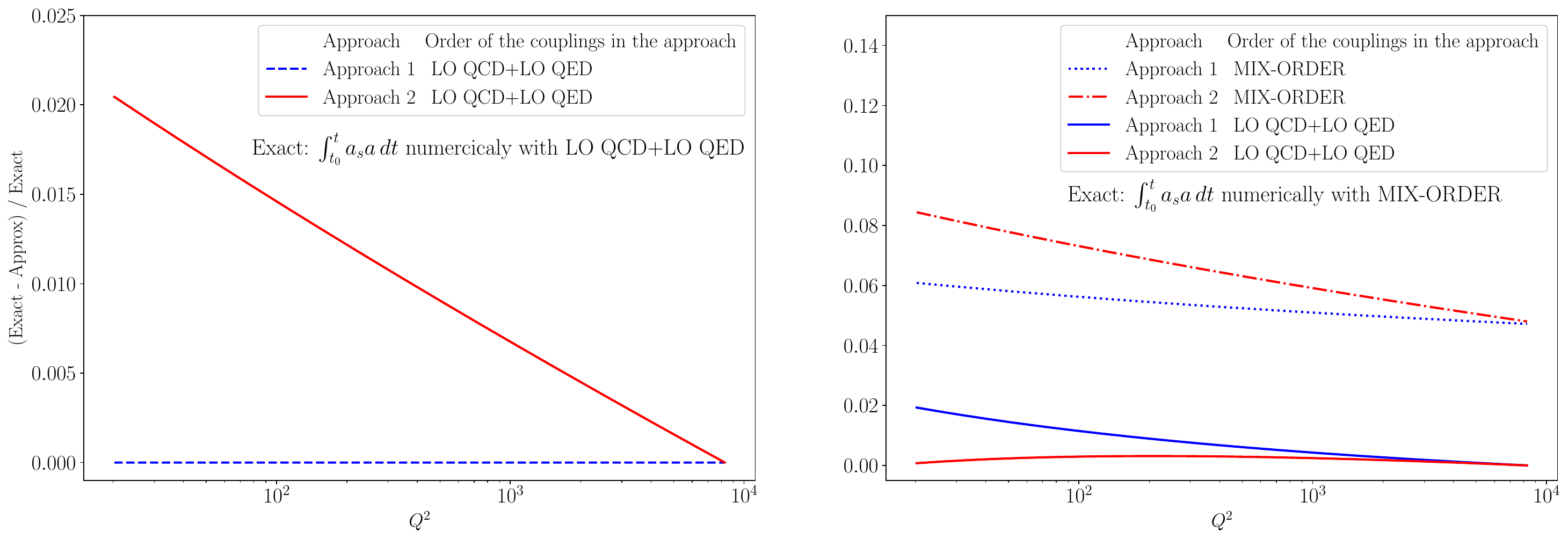}
    \caption{Integral $\int_{t_0}^t \as\al \, dt$ with $Q_0^2 = M_z^2$. The curves show the relative difference with respect to the exact numerical solution, for the approximations given by Eq.(\ref{eq:int11-best}) (blue) and Eq.(\ref{eq:int113}) (red). The left plot corresponds to the exact solution using the LO approximation for both couplings, while the right plot includes the mixed-order evolution.}
    \label{fig:int11numvsana}
\end{figure}

We propose two alternatives to compute the mixed-order integral of Eq.(\ref{eq:Eintegral}). 
The first one uses the LO approximation for both couplings, $\as$ and $\al$, since higher-order terms in the RGE, Eq.(\ref{eq:coupling}), would go beyond our intended accuracy. The evolution operators for the LO couplings of QCD and QED are given by Eq.(\ref{eq:couplingevoloperadoresQCD}) and Eq.(\ref{eq:couplingevoloperadoresQED}) respectively, setting $\beta_{20} \rightarrow0$. Replacing in Eq.(\ref{eq:Eintegral}) and performing the integral we obtain,
\beqn
\ln{E^{\text{MIX}}}(Q,Q_0) &=& R^{(1,1)} \int_{t_0}^t \as(t)\, \al(t) \, dt \nn \\ 
&=&  R^{(1,1)}
\frac{\aso\al_0 }{ \aso \beta_{10} -\al_0 \beta'_{01} }\left[ \ln{1+\aso  \beta_{10} \Delta t} - \ln{1+\al_0  \beta'_{01} \Delta t}
\right] \nn\\
&=&  R^{(1,1)}
\frac{\aso\al_0}{ \aso \beta_{10} -\al_0 \beta'_{01} } \ln {\frac{\aso \al(t)}{\as(t)\al_0}}.
 \label{eq:int11-best}
 \eeqn
The second approach is the following: define a {\it mixed coupling} $\gamma=\as \al$ whose RGE can be easily obtained as
\beqn
 \frac{d\gamma}{dt}=\as \frac{d\al}{dt} + \al \frac{d\as}{dt} \simeq  -\gamma^2 \left(\frac{\beta'_{01} \al +\beta_{10} \as}{\as \al}\right),
 \label{eq:gamma2}
 \eeqn
where we again use the LO approximation for both couplings.
Now, to the accuracy we are interested, we might attempt to consider the term in brackets as {\it constant}, in the sense that for the integration we can replace  $\as(t) \rightarrow \aso$ and $\al(t) \rightarrow \al_0$. In that way, Eq.(\ref{eq:Eintegral}) results in
\beqn
\ln{E^{\text{MIX}}}(Q,Q_0)&=& R^{(1,1)} \int_{\gamma(t_0)}^{\gamma(t)} \gamma(t) \, \frac{dt}{d\gamma} d\gamma = - R^{(1,1)} \left(\frac{\aso \al_0}{\beta'_{01} \al_0 +\beta_{10} \aso}\right) \ln{\frac{\gamma(t)}{\gamma(t_0)}}\nn \\
 &=&- R^{(1,1)}\left(\frac{\aso \al_0}{\beta'_{01} \al_0 +\beta_{10} \aso}\right)  \left[ \ln{\frac{\as(t)}{\aso}} + \ln{\frac{\al(t)}{\al_0}} \right]
 \nn \\
  &=&
  - R^{(1,1)}\left(\frac{\aso \al_0}{\beta'_{01} \al_0 +\beta_{10} \aso}\right)   \ln{\frac{\as(t) \al(t)}{\aso\al_0}}.
  \label{eq:int113}
 \eeqn
Notice that we can choose to employ a different perturbative order for the couplings in Eqs.(\ref{eq:int11-best}) and (\ref{eq:int113}), expecting that such a choice would lead to corrections beyond the accuracy we are working with. To compare both approaches, in Fig.\ref{fig:int11numvsana} we present the relative difference between the integral in Eq.(\ref{eq:Eintegral}), computed using the approach 1 (Eq.(\ref{eq:int11-best}), blue curves) and approach 2 (Eq.(\ref{eq:int113}), red curves), with respect to the exact result obtained by numerical integration.

In the left plot, both approaches are computed using the LO approximation for the couplings and compared to the exact solution computed at the same accuracy (i.e., using the LO approximation for both couplings in the numerical integration). At this order, approach 1 is identical to the exact result, as expected, since it solves the integral using the LO couplings. On the other hand, approach 2 exhibits deviations of approximately $\sim 1\%$.

In the right plot, both approaches are compared to the exact solution computed up to mixed order, that is, using Eqs.~(\ref{eq:couplingevoloperadoresQCD}) and (\ref{eq:couplingevoloperadoresQED}) for the coupling evolution in the numerical integration. The solid lines represent both approaches computed using the LO approximation for the couplings (following the same colour code as before), while the dashed lines correspond to the same computation but employing the mixed-order approximation for the interactions, Eq.(\ref{eq:couplingevoloperadoresQCD}) and Eq.(\ref{eq:couplingevoloperadoresQED}). The relative differences with respect to the exact mixed-order solution are slightly larger in the latter case than those obtained with the LO approximation; however, the discrepancies remain at the percent level. Note that these percent level differences refer to corrections that are already of percent size or smaller. Therefore, the observed discrepancies correspond to higher-order effects that lie beyond the perturbative accuracy considered in this work. In the remainder of the paper, we will use approach 1 with the mixed-order evolution for the coupling.
\subsection{Singlet case}
For the singlet sector, we can express the evolution equations system Eqs. (\ref{eq:evolucionDUDSIMPLE},
\ref{eq:evolucionSIGMASIMPLE},
\ref{eq:evolucionGLUONSIMPLE} and
\ref{eq:evolucionGAMMASIMPLE}) in a matrix form,
\beqn
\frac{d \bar{f}^S(Q)}{d t}= \m{P} . \bar{f}^S(Q),
\label{eq:singlete}
\eeqn
where $\bar{f^S}$ is the vector of singlet combinations of flavours $\bar{f^{S}}=\{\Delta_{UD}, \Sigma,g,\gamma\}$ ,
and $\m{P}$ is the matrix of kernels, which can also be expressed in a perturbative form, up to the order $\mathcal{O}(\as\, \al)$ reads,
\beqn 
\m{P}=\as \m{P}^{(1,0)}+\as^2 \m{P}^{(2,0)}+\al \m{P}^{(0,1)} + \as \al \m{P}^{(1,1)} , 
\label{eq:Pmatriz}
\eeqn
where $\m{P}^{(1,0)}$,$\m{P}^{(2,0)}$ and $\m{P}^{(0,1)}$ can be found in \cite{deFlorian:2023zkc}, and
\beqn
\m{P}^{(1,1)}=\left(\begin{array}{llll}
\eta ^+ P_{f}^{+(1,1)}& \eta ^- P_{f}^{+(1,1)} & \delta^2 P_{fg}^{(1,1)} & \delta^2 P_{f\gamma}^{(1,1)}\\
\eta ^- P_{f}^{+(1,1)} & \eta ^+P_{f}^{+(1,1)} & e^2_\Sigma P_{fg}^{(1,1)} & e^2_\Sigma P_{f\gamma}^{(1,1)}\\
\eta ^- P_{g f}^{(1,1)} & \eta ^+ P_{g f}^{(1,1)} & P_{gg}^{(1,1)} &  P_{g\gamma}^{(1,1)}\\
\eta ^- P_{\gamma f}^{(1,1)}& \eta ^+ P_{\gamma f}^{(1,1)} &P_{\gamma g}^{(1,1)} & P_{\gamma\gamma}^{(1,1)}
\end{array}\right),
\label{Eq:Pqed}
\eeqn
where $P_{f}^{+(1,1)}$,$P_{fg}^{(1,1)}$,$P_{f\gamma}^{(1,1)}$,$P_{gf}^{(1,1)}$ and $P_{\gamma f}^{(1,1)}$, can be extracted from the definitions,
\beqn
P_{u(d)\gamma}^{(1,1)}&=&e_{u(d)}^2 P_{f\gamma}^{(1,1)},\nn\\
P_{u(d)g}^{(1,1)}&=&e_{u(d)}^2 P_{fg}^{(1,1)},\\
P_{g u(d)}^{(1,1)}&=&e_{u(d)}^2 P_{g f}^{(1,1)}\nn,\\
P_{\gamma u(d)}^{(1,1)}&=&e_{u(d)}^2 P_{\gamma f}^{(1,1)},\nn\\
P_{u(d)}^{+(1,1)}&=&e_{u(d)}^{2}P_{f}^{+(1,1)},\nn
\eeqn
using Eqs.(\ref{eq:kernels1}, \ref{eq:kernels4}, \ref{eq:kernels7}, \ref{eq:kernels8}, \ref{eq:kernels10}, \ref{eq:kernels11}) (with the notation $q=u,d$) for the kernels in the first column, respectively. In addition, we define,
\beqn
\eta^{\pm}&=&\frac{1}{2}\, (e_u^2 \pm e_d^2),  \\
e^2_\Sigma&=&2\, (n_u \, e_u^2+n_d\, e_d^2),\\ 
\delta^2&=&2\, (n_u \, e_u^2-n_d\, e_d^2).
 \eeqn
Proposing a solution of the form $\bar{f}^S(Q)=\m{E}(Q,Q_0)\bar{f}^S(Q_0)$, we obtain the following matrix linear equation analogous to Eq.(\ref{eq:evolnonsinglet}) for the evolution operator,
\beqn
\frac{d\m{E}(Q,Q_0)}{dt}=(\as \m{P}^{(1,0)}+\as^2 \m{P}^{(2,0)}+\al \m{P}^{(0,1)} + \as \al \m{P}^{(1,1)})\m{E}(Q,Q_0).
\label{eq:evolnloSinglet}
\eeqn
Because of the non-commutativity between the $\m{P}^{(i,j)}$ matrices, the singlet case is more complicated than the non-singlet one. 
We present two ways to solve these linear differential equation systems. The first method is a modification of the $U$-matrix method \cite{Vogt:2004ns,deFlorian:2023zkc} that incorporates mixed-order corrections. It is an extension of the non-singlet case to matrix linear equations. The key observation is that we have mixed-order contributions not only due to the inclusion of the AP kernels but also because of the evolution of the coupling, as discussed in the previous section. Similarly to the non-singlet case, the approach involves reusing the QCD and QED operators that have already been established \cite{deFlorian:2023zkc,Vogt:2004ns,Gluck:1989ze} and defining a new operator that incorporates all contributions of mixed-order. This is advantageous because only the QCD $\otimes$ QED operator needs to be computed, and from an implementation perspective, it allows for the straightforward reuse of legacy QCD codes (e.g., \cite{deFlorian:2023zkc,Vogt:2004ns}).
For the second method, we introduce the Magnus expansion approach. A recent work \cite{Simonelli:2024vyh} proposed this method to solve the NLO QCD PDF evolution. One of the most important advantages is that it allows one to obtain a general solution with a closed exponential form for any linear ordinary differential equation. However, the solution becomes complicated to manage at higher orders, involving more complex integrals compared to the $U$-matrix method, which more easily allows the reuse of existing QCD and QED evolution codes.

\subsubsection*{Method 1: $\m{E}^{\text{MIX}}$ with mixed-order coupling evolution}
\label{sec:solution}
As in the non-singlet case, we reuse the available operators from pure QCD and QED evolution~\cite{deFlorian:2023zkc,Vogt:2004ns}. We then introduce an operator, $\m{E}^{\text{MIX}}$, which accounts for all QCD~$\otimes$~QED effects, including the evolution of the couplings. Given that the QCD and QED kernels do not commute and in principle it is not possible to solve the singlet evolution analytically in a closed form \cite{NNPDF:2024djq}, we introduce the sum of operators $\m{A}_i$ to correct for the non-commutative behaviour of the matrix kernels. 

The solution takes the form
\beqn
\m{E}^{\text{U-mat}}=\left(1+\sum_{i=0}^\infty\m{A}_i\right)\m{E}^{\text{MIX}}\m{E}^{\text{QED}}\m{E}^{\text{QCD}},
\label{eq:solumix}
\eeqn
where $\m{E}^{\text{QCD}}$ and $\m{E}^{\text{QED}}$ are the solutions for the pure QCD and QED cases, respectively~\cite{deFlorian:2023zkc}. The differential equations for \( \m{E}^{\text{QED}} \) and \( \m{E}^{\text{QCD}} \), keeping terms up to \( \mathcal{O}(\as \al) \), are given by
\beqn
\frac{d \m{E}^{\text{QCD}} }{dt}&=& \frac{d \m{E}^{\text{QCD}}}{d\as} \frac{d\as}{dt}=  \as  
( \m{P}^{(1,0)} + \as \m{P}^{(2,0)} + \al \frac{\beta_{11}}{\beta_{10}} \m{P}^{(1,0)}) \m{E}^{\text{QCD}} \equiv \m{S}_{\text{QCD}}\m{E}^{\text{QCD}} \, ,
\label{eq:QCDderitS}
\eeqn
and
\beqn
\frac{d \m{E}^{\text{QED}} }{dt}&=& \frac{d \m{E}^{\text{QED}} }{d\al} \frac{d\al}{dt}=  \al \m{P}^{(0,1)} 
( 1 +  \as \frac{ \beta'_{11}}{\beta'_{01}}) \m{E}^{\text{QED}}\equiv \m{S}_{\text{QED}}\m{E}^{\text{QED}},
\label{eq:QEDderitS}
\eeqn
where Eq.~(\ref{eq:coupling}) has been used to express the derivatives of the couplings with respect to $t$.

For the $\m{E}^{\text{MIX}}$ operator we propose,
\beqn
\frac{d \m{E}^{\text{MIX}}}{dt}=  \as \, \al\,  \left( \m{P}^{(1,1)} -\frac{\beta_{11}}{\beta_{10}} \m{P}^{(1,0)} 
-\frac{\beta'_{11}}{\beta'_{01}}  \m{P}^{(0,1)} \right) \m{E}^{\text{MIX}} \equiv   \as \, \al\,   \m{R}^{(1,1)}  \m{E}^{\text{MIX}},
\label{eq:derivadaEmixS}
\eeqn
which implies that $\m{E}^{\text{MIX}}$ takes the form
\beqn
\m{E}^{\text{MIX}}=e^{\m{R}^{(1,1)} \int_{t_0}^{t}  \,\as \al \,dt},
\label{eq:Emix}
\eeqn
where the integral can be computed using the same approach as in Eq.~(\ref{eq:int11-best}) or Eq.~(\ref{eq:int113}), or numerically. As we said before, the $\m{A}_i$ operators in Eq.~(\ref{eq:solumix}) are responsible for cancelling terms that arise due to the non-commutative nature of the kernels matrices. By differentiating Eq.~(\ref{eq:solumix}) using Eqs.~(\ref{eq:QCDderitS}, \ref{eq:QEDderitS}, \ref{eq:derivadaEmixS}), and then comparing the result with Eq.~(\ref{eq:evolnloSinglet}), we obtain the following expression for each $\m{A}_i$ with $i>0$,
\beqn
\frac{d\m{A}_i}{d t}&=& \m{A}_{i-1} \frac{d\m{A}_0}{d t}- \left[\m{A}_{i-1},\as \m{P}^{(1,0)}+\as^2 \m{P}^{(2,0)}+\al \m{P}^{(0,1)}+\as\al \m{P}^{(1,1)}\right], \,\,\,\,
\label{eq:Ai}
\eeqn
which expresses the $i$-th term, $\m{A}_i$, in terms of $\m{A}_{i-1}$, allowing for an iterative solution. The first operator, $\m{A}_0$, must satisfy the following equation in order to properly correct the spurious non-commutative contributions that would otherwise spoil the solution \footnote{To obtain this expression, we use that $\m{E}^{\text{QED}}$ and $\m{E}^{\text{MIX}}$ have simple exponential forms, $e^{\m{I}_{\text{QED}}(t,t_0)}$ and $e^{\m{I}_{\text{MIX}}(t,t_0)}$, respectively. However, the derivation of $\m{A}_0$ can be performed with any type of expression for these operators, even when going to higher orders in QED.},
\beqn
\frac{d\m{A}_0}{d t}&=&  -\sum_{n=1}^\infty \frac{1}{n!} [\m{I}_{\text{QED}}(t,t_0), \m{S}_{\text{QCD}}(t)]_n -\sum_{n=1}^\infty \frac{1}{n!} [\m{I}_{\text{MIX}}(t,t_0), \m{S}_{\text{QED}}(t)]_n \nn\\
&-&\sum_{n=1}^\infty \frac{1}{n!} [\m{I}_{\text{MIX}}(t,t_0),\m{S}_{\text{QCD}}(t)+ \sum_{n=1}^\infty \frac{1}{n!} [\m{I}_{\text{QED}}(t,t_0), \m{S}_{\text{QCD}}(t)]_n]_n, \,\,\,\,\,\,
\label{eq:A0}
\eeqn 
where we define \( \m{I}_{\text{QED}}(t_1,t_2) = - \ln{\frac{\al(t_1)}{\al(t_2)}} \frac{\m{P}^{(0,1)}}{\beta^{\prime}_{01}} \), $\m{I}_{\text{MIX}}(t_1,t_2) = \m{R}^{(1,1)} \int^{t_1}_{t_2} \as \al\,dt$. Also, we define, $[A, B]_n = [A, [A, \dots, [A, B]]]$, where the commutator is nested $n$ times. Using the identity \( \sum_{n=1}^{\infty}\frac{1}{n!}[A, B]_n = e^A B e^{-A}-B \), valid for any matrices \( A \) and \( B \), we can compute \( \m{A}_0 \) and $\frac{d\m{A}_0}{dt}$ analytically to all orders in \( n \). 

Solving the full Eq.~(\ref{eq:Ai}) would provide the solution to the problem of the non-commutative behaviour between the QCD and QED matrix kernels. 
However, no analytical solution can be obtained for the full expression, nor is it possible to expand the solution as a pure power series in the coupling constants. Nevertheless, up to mixed order, it is numerically sufficient to retain only the dominant contribution in Eq.~(\ref{eq:Ai}), namely the commutator $[\m{A}_{i-1},\, \as \m{P}^{(1,0)}]$.
Proceeding in this way, we can resume the contribution of the this term. The sum \( \sum_{i=0}^{\infty} \m{A}_i \) is then expressed as,
\beqn
\sum_{i=0}^\infty \m{A}_i (t)&=&  \int_{t_0}^t e^{\m{I}_{\text{QCD}}(t,t_1)} \,\frac{d\m{A}_0}{dt}(t_1)\, e^{-\m{I}_{\text{QCD}}(t,t_1)}\, dt_1 , \,\,\,\,\,\,\,\,\,\,\,\,
\label{eq:AiLOfinal}
\eeqn 
where we define $\m{I}_{\text{QCD}}(t_1,t_2) = - \ln{\frac{\as(t_1)}{\as(t_2)}} \frac{\m{P}^{(1,0)}}{\beta_{10}} $. To obtain Eq.(\ref{eq:AiLOfinal}), we use the LO approximation for the couplings.

To simplify the analytical computation, we can retain only the dominant term in Eq.~(\ref{eq:A0}), i.e., $\sum_{n=1}^\infty \frac{\as}{n!} [\m{I}_{\text{QED}}(t,t_0), \m{P}^{(1,0)}]_n$, as the remaining terms yield very small contributions. Although this approximation is not really mandatory, since $\frac{d\m{A}_0}{dt}$ can be fully resummed using the same method described below Eq.~(\ref{eq:A0}), it significantly simplifies the analytical calculation. We obtain
\beqn
\frac{d\m{A}_0}{dt}(t) = \as(t) \left( \m{P}^{(1,0)} - e^{\m{I}_{\text{QED}}(t,t_0)} \, \m{P}^{(1,0)} \, e^{-\m{I}_{\text{QED}}(t,t_0)} \right).
\label{eq:A0simple}
\eeqn
By diagonalizing both \( \m{I}_{\text{QED}} \) and \(\m{I}_{\text{QCD}} \), we can compute the matrix products in Eqs.~(\ref{eq:A0simple}) and~(\ref{eq:AiLOfinal}), obtaining explicit matrix expressions for \( \frac{d\m{A}_0}{dt} \) and \( \sum_{i=0}^\infty \m{A}_i \). For implementation purposes, we evaluate Eq.(\ref{eq:AiLOfinal}) numerically, using Eq.(\ref{eq:A0simple}) for $\frac{d\m{A}_0}{dt}$.
\subsubsection*{Method 2: Exponential solution, ``Magnus'' expansion}
As we explained before, the advantage of the Magnus method is that it allows us to write the solution in a closed exponential form. However, as we will see, going to higher orders can be challenging and the problem of non-commuting operators still remains. A disadvantage is that with this method, we do not reuse the well-known solutions for pure QCD and pure QED; instead, we have to solve the full differential equation from the beginning.
The Magnus method postulates that for a linear ordinary differential equation of the form,
\beqn
\frac{d \m{E}^{\text{Mag}}(t)}{dt}=\m{A}(t)\, \m{E}^{\text{Mag}}(t),
\eeqn
the general solution takes the form,
\beqn
\m{E}^{\text{Mag}}(t)=e^{\Omega(t)} \,\,\,\, \text{with,   } \Omega(t)=\sum_i^\infty \Omega_i(t),
\label{eq:magsolu}
\eeqn
where the first two terms of $\Omega$ are
\beqn
\Omega_{1}(t)  &=&\int_{t_0}^{t} dt_{1} \m{A}(t_{1}),\nonumber\\\Omega_{2}(t)  & =&\frac{1}{2}\int_{t_0}^{t}dt_{1}\int_{t_0}^{t_{1}} dt_{2}\ \left[  \m{A}(t_{1}),\m{A}(t_{2})\right].\label{eq:omegamagnus}
\eeqn
In the DGLAP equations with mixed-order corrections $\m{A}(t)$ is given by Eq.(\ref{eq:evolnloSinglet}), we have $\m{A}(t)=\as \m{P}^{(1,0)}+\as^2 \m{P}^{(2,0)}+\al \m{P}^{(0,1)}+ \as \al \m{P}^{(1,1)}$. 
Keeping the leading term in $\Omega_2$, we find that
\beqn
\m{E}^{\text{Mag}}&=&\exp \left[ \int^t_{t_0} \, dt_1\, (\as(t_1) \m{P}^{(1,0)}+\al(t_1) \m{P}^{(0,1)}+\as^2(t_1) \m{P}^{(2,0)}+\as(t_1) \al(t_1) \m{P}^{(1,1)} )\nn \right.\\ 
&+&\left.\frac{1}{2}\int^{t}_{t_0}dt_1 \int^{t_1}_{t_0}dt_2 \left(\as(t_1) \al(t_2)-\as(t_2) \al(t_1)\right) [\m{P}^{(1,0)},\m{P}^{(0,1)}] \right].
\label{eq:solumagnus}
\eeqn
To increase the accuracy of the solution, one needs to add terms to the Magnus expansion. However, the expressions become difficult to handle for $\Omega_i$ with \(i > 3\) \cite{Simonelli:2024vyh,Blanes:2008xlr}. 
For the first two integrals in Eq.(\ref{eq:solumagnus}) we use the full RGE evolution equations for the couplings, as given in Eq.(\ref{eq:coupling}), we obtain
\beqn
 \int^t_{t_0} \as(t_1) \, dt_1 &\simeq& -\frac{1}{\beta_{10}}\ln{\frac{\as(t)}{\aso}}+\frac{\beta_{20}}{\beta^2_{10}}(\as(t)-\aso)-\frac{\beta_{11}}{\beta_{10}}\int_{t_0}^t \as(t_1) \al(t_1) \, dt_1 ,
 \label{eq:intas}
\eeqn
\beqn
 \int^t_{t_0} \al(t_1) \, dt_1 &\simeq& -\frac{1}{\beta^\prime_{01}}\ln{\frac{\al(t)}{\alo}}-\frac{\beta^\prime_{11}}{\beta^\prime_{01}}\int_{t_0}^t \as(t_1) \al(t_1)\, dt_1 ,
 \label{eq:intal}
\eeqn
where we keep contributions up to mixed-order accuracy. For the third term, we employ the LO QCD approximation for the coupling, excluding the corresponding higher-order QCD and mixed-order corrections, as these would introduce contributions beyond the intended accuracy,
\beqn
\int^t_{t_0} \as^2(t_1) dt_1&\simeq&-\frac{1}{\beta_{10}}(\as(t)-\aso).
\eeqn
For the mixed-order integral, $\int \as \al \, dt$, in Eqs (\ref{eq:solumagnus}), (\ref{eq:intas}) and (\ref{eq:intal}), we employ the results from Sec. \ref{sec:nonsinglet}. Finally, for the last term in Eq.(\ref{eq:solumagnus}), we utilise the LO QED and LO QCD couplings,
\beqn
\int_{t_0}^{t} &dt_1& \int_{t_0}^{t_1} dt_2(\as(t_1) \al(t_2) - \as(t_2) \al(t_1)) =\frac{1}{2 \beta_{10} \beta'_{01}} \times
  \nn  \\  &\bigg \{&
    - \left. \ln{1 + \aso \beta_{10} \Delta t} \ln{
    \frac{\aso \beta_{10} (1 + \alo \beta'_{01} \Delta t)}{\aso \beta_{10} - \alo \beta'_{01}}
    }- \operatorname{Li}_2\left(
    \frac{\aso \beta_{10}}{\aso \beta_{10} - \alo \beta'_{01}}
    \right) \right. \notag \\
    & +& \left. \ln{1 + \alo \beta'_{01} \Delta t} \ln{
    \frac{\alo (\beta'_{01} + \aso \beta_{10} \beta'_{01} \Delta t)}{-\aso \beta_{10} + \alo \beta'_{01}}
    }
    + \operatorname{Li}_2\left(
    \frac{\alo \beta'_{01}}{-\aso \beta_{10} + \alo \beta'_{01}}
    \right) \right.  \\
    & -&  \operatorname{Li}_2\left(
    \frac{\alo \beta'_{01} (1 + \aso \beta_{10} \Delta t)}{-\aso \beta_{10} + \alo \beta'_{01}}
    \right)
    + \operatorname{Li}_2\left(
    \frac{\aso \beta_{10} (1 + \alo \beta'_{01} \Delta t)}{\aso \beta_{10} - \alo \beta'_{01}}
    \right) \bigg\}, \nn
\eeqn
where $\operatorname{Li}_2$ is the polylogarithm function.
With all these ingredients, we can perform the evolution of the PDFs using the operator $\m{E}^{\text{Mag}}$. It should be noted that the operator $\m{E}^{\text{Mag}}$ satisfies Eq.(\ref{eq:evolnloSinglet}) up to spurious terms arising from the non-commutative behaviour, i.e., the non-commutative issue is still present at the time of providing a formally ordered perturbative expansion. In principle, including higher-order contributions to $\Omega$ would lead to a more accurate solution, although, as mentioned before, the calculation of such higher-order terms rapidly becomes increasingly complicated. Nonetheless, phenomenological applications and comparisons with other models are still feasible.


For implementation purposes, it is convenient to express the operators 
$\m{E}^{\text{MIX}}$ in Eq.~(\ref{eq:Emix}) and 
$\m{E}^{\text{Mag}}$ in Eq.~(\ref{eq:solumagnus}) 
in terms of the projectors 
$\m{e}^{\text{U}}_i$ and $\m{e}^{\text{Mag}}_i$, respectively, as follows,
\beqn
\m{E}^{\text{Mag/MIX}}=\m{e}^{\text{Mag/U}}_1e^{\lambda^{\text{Mag/U}}_1 }+\m{e}^{\text{Mag/U}}_2e^{\lambda^{\text{Mag/U}}_2 }+\m{e}^{\text{Mag/U}}_3e^{\lambda^{\text{Mag/U}}_3 }+\m{e}^{\text{Mag/U}}_4e^{\lambda^{\text{Mag/U}}_4},
\label{eq:descomposicionE}
\eeqn
where the procedure for determining the eigenvalues $\lambda_i^{\text{Mag/U}}$ and the projectors $\m{e}^{\text{Mag/U}}_i$ is detailed in Appendix.\ref{sec:MatrixR}.

In Fig.\ref{fig:pdf} we show the pPDFs mixed-order relative corrections to QCD+QED for the polarized case at $Q^2=1000 \gev$, defined as,
\beqn
\delta f= \frac{\Delta f^{\textit{QCD+QED}}-\Delta f^{\textit{QCD+QED+MIX}}}{\Delta f^{\textit{QCD+QED}}},
\label{eq:relativef}
\eeqn
where $\Delta f^{\textit{QCD+QED}}$ refers to the parton densities evolved with NLO QCD + LO QED corrections, and $\Delta f^{\textit{QCD+QED+MIX}}$ includes, in addition, the mixed-order contributions.
We show the results for both the $U$-matrix(blue dashed line) method and for the ``Magnus expansion"(red line) method. We used the DSSV pPDF set \cite{Borsa:2024mss} for the initial scale $Q_0^2=1 \gev$, and for the polarized photon density, we used the results of \cite{deFlorian:2024hsu}. We use a variable flavour number scheme with a minimum of three flavours
and set the charm and bottom threshold at $m^2_c = 2\gev$ and $m^2_b = 20.25 \gev$ respectively
(we do not take into account the top quark.). First, we observe that both methods predict similar mixed‑order relative corrections across all pPDF combinations. In the case of the non‑singlet combination, as $\Delta u_v$ (lower left panel), the predictions coincide exactly, since these distributions are treated identically in both methods (and are decoupled from the singlet combinations). Relative corrections are approximately of order $\mathcal{O}(10^{-4})$ for almost all pPDFs, with the exception of the photon pPDF, for which they become significantly larger, reaching the percent level. Even at high values of $x$, where the photon pPDF approaches zero, the corrections reach up to $\sim 12\%$. Such mixed‑order contributions to the photon pPDF may be particularly relevant for processes sensitive to it, for example, single‑inclusive prompt photon production in electron–proton collisions \cite{Rein:2024fns}.

We also compute the relative difference between the Magnus and $U$-matrix methods, which arises mainly from the different treatment of higher‑order QCD terms between methods. This discrepancy exceeds the mixed-order corrections for almost all pPDFs combinations, except for the photon pPDF, for which the mixed-order corrections are larger than the other pPDFs, as previously observed, and for the non-singlet combination, since, as mentioned earlier, both methods are equivalent in this case. The fact that higher-order QCD corrections are larger than mixed-order corrections can be attributed to the dominant impact of the evolution occurring at low values of $Q^2$, where the strong coupling constant is significantly larger. However, one could include higher-order QCD contributions in both the Magnus method, Eq.(\ref{eq:magsolu}) and the $U$-matrix method, i.e. $\m{E}^{\text{QCD}}$ of Eq.(\ref{eq:solumix}) (i.e. using higher-order terms in Eq.(2.24) from \cite{Vogt:2004ns}), to reduce this gap.

Considering practical applicability, although the Magnus expansion provides a closed-form solution, it involves more complex integrals compared to the $U$-matrix method, which more readily reuses existing QCD and QED evolution codes. Therefore, in the remainder of the manuscript, we rely on the conventional $U$-matrix method.

\begin{figure}
   \centering
    \includegraphics[width=0.95\textwidth]{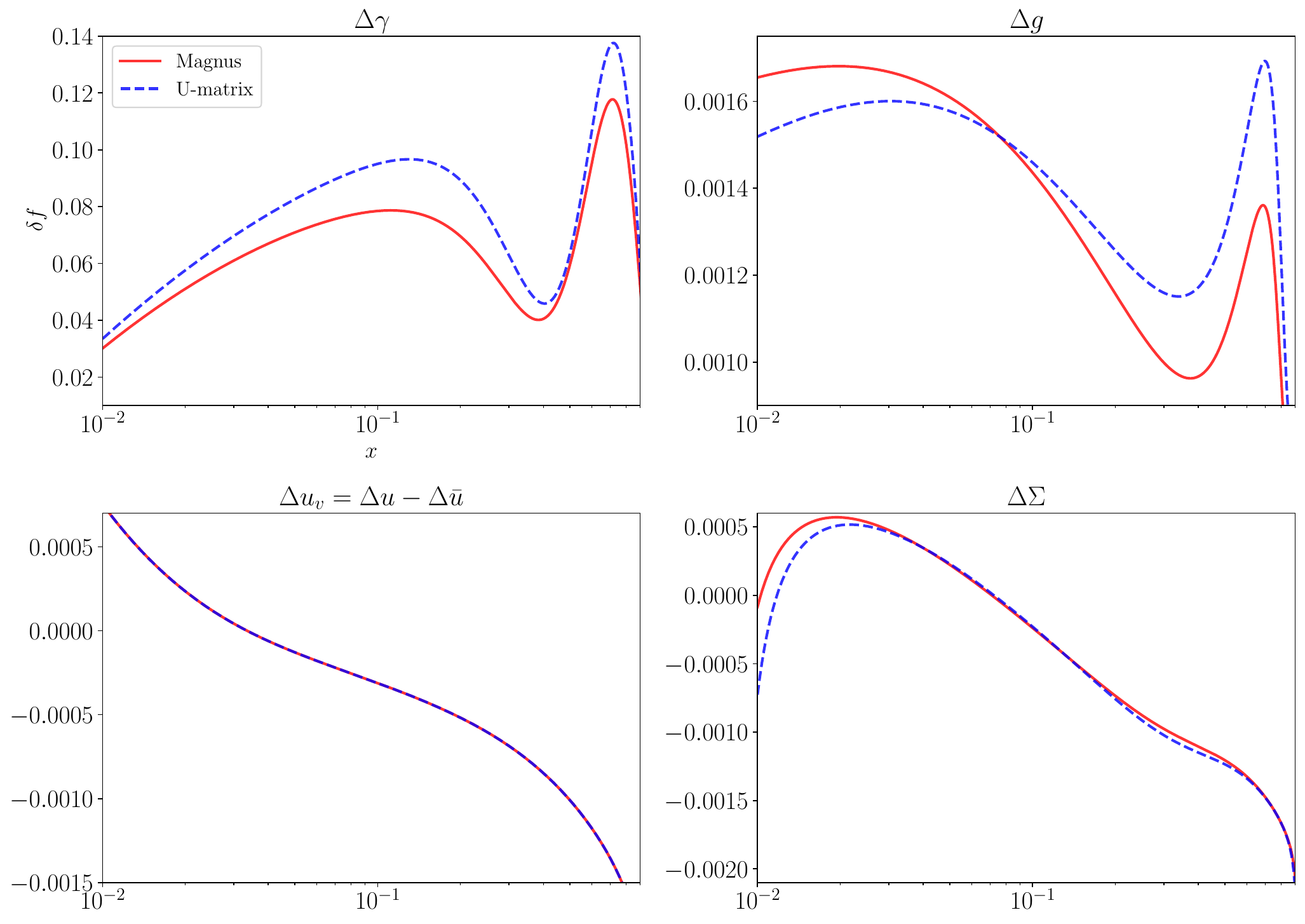}
    \caption{Relative corrections to PDFs in the polarized case, at $Q^2=1000 \gev$, for the Magnus  approach (red) and the $U$-matrix approach (blue dashed line).}
    \label{fig:pdf}
\end{figure}

\section{$g_1$ at mixed-order}\label{sec:g1}
\begin{figure}
  \centering
   \includegraphics[width=0.95\textwidth]{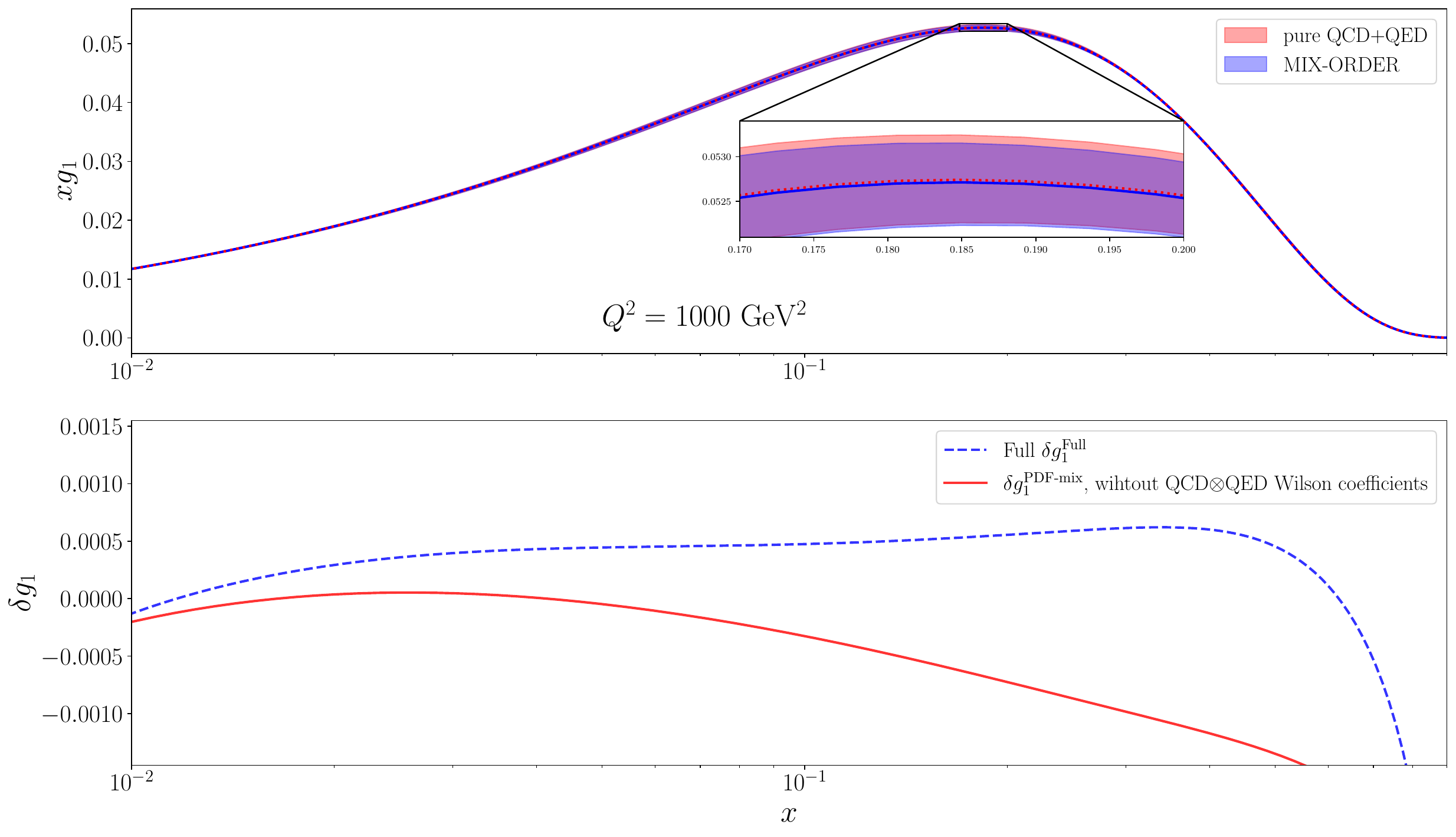}
   \caption{Structure function $g_1$ at $Q^2 = 1000~\gev$ (top plot). The red dashed line includes pure QCD and QED corrections, while the blue line incorporates mixed-order corrections. The bands correspond to the theoretical uncertainty obtained by the independent variation of the renormalization and factorization scales. The bottom plot shows the relative corrections, as defined in Eq.(\ref{eq:relativef}), for the full computation (blue dashed line) and for the computation with the mixed-order Wilson coefficients subtracted (red solid line).}\label{fig:g1}
\end{figure}

In order to estimate the effect of mixed-order corrections directly on physical observables, we compute the mixed-order corrections for the polarized structure function $g_1$. Although $g_1$ with QED corrections must include both photon radiative and two-photon exchange contributions \cite{Cammarota:2025jyr,Liu:2020rvc}, we assume, in order to isolate the impact of our mixed-order corrections, that these effects can be treated independently, as it is customary. At leading twist, the structure function can be expressed through a perturbative expansion in terms of the Wilson coefficient $\Delta C_i^{(i,j)}$, where $i$ labels the different parton channels. Taking into account the quark, gluon, and photon channels, it is given by,
\beqn
g_1(x,Q^2)&=&\frac{1}{2}\sum_q e_q^2 \{(\Delta q+\Delta \bar{q})+ 2\,\as\,[\Delta C_q^{(1,0)} \otimes (\Delta q+\Delta \bar{q})+ \Delta C_g^{(1,0)} \otimes\Delta g]\nn 
 \\
&+&2\,\al\,\left[\Delta C_q^{(0,1)}\otimes(\Delta q+\Delta \bar{q})+\Delta C_\gamma^{(0,1)}\otimes \Delta \gamma \right] \label{eq:g1mix}\\
  \,&+&4\,\as\al  \left[\Delta C_q^{(1,1)}\otimes(\Delta q+\Delta \bar{q})+\Delta C_g^{(1,1)}\otimes \Delta g+\Delta C_\gamma^{(1,1)}\otimes \Delta \gamma \right]
\},\nn 
\eeqn
where the sum runs over all quark flavours. We performed all the computations in the $\overline{\text{MS}}$ scheme. The QCD coefficients, $\Delta C^{(1,0)} _i$, can be found in \cite{Zijlstra:1993sh}. The QED coefficients $\Delta C^{(0,1)}_i$ were calculated in \cite{deFlorian:2023zkc}. As we have already mentioned, for the calculation of the mixed-order contributions $\Delta C^{(1,1)}_i$, we apply the ``Abelianization'' technique to the NNLO coefficients, which can be found in \cite{Zijlstra:1993sh}. The results are shown in Appendix.\ref{sec:apB}, expressed in the $x$-space.
Then with the MT-1.0 Mathematica package \cite{Hoschele:2013pvt} we transform all coefficients to the Mellin $N$-space. Using the properties presented in \cite{Blumlein:2009ta,Blumlein:1998if} for special functions, we develop a FORTRAN code that computes the structure-function $g_1$ to mixed-order. 
Figure~\ref{fig:g1} shows $g_1$ evaluated at $Q^2 = 1000~\gev$, including pure QCD and QED corrections (red dashed line) and the full result with mixed-order corrections (blue line). The bands at each order represent the theoretical uncertainty of the cross section, obtained from the independent 7-point variation of the renormalization and factorization scales ($Q/2 \le \mu_{F}, \mu_{R} \le 2Q$ with $1/2 \le \mu_{F}/\mu_{R} \le 2$). We observe that, once the QCD$\otimes$QED contributions are included, the scale uncertainty decreases by an amount comparable to the size of the mixed-order effects themselves. However, the resulting scale uncertainty remains significantly larger than the mixed-order correction. The evolution of the pPDFs is performed using the $U$-matrix approach. The lower panel of Fig.\ref{fig:g1} displays the relative mixed-order corrections, as defined in Eq.(\ref{eq:relativef}) (blue dashed line). The relative corrections are found to be of the order $\mathcal{O}(10^{-4})$ for $x \simeq 0.1$, and increase to $\mathcal{O}(10^{-3})$ for $x \simeq 1$, where $g_1$ approach to zero. In order to isolate the mixed-order contributions of the Wilson coefficients from those arising from pPDF evolution, we define
\beqn
g_1^{\text{PDF-mix}}=g_1^{\text{Full}}-
  2\,\as\al  [\Delta C_q^{(1,1)}\otimes(\Delta q+\Delta \bar{q})+\Delta C_g^{(1,1)}\otimes \Delta g+\Delta C_\gamma^{(1,1)}\otimes \Delta \gamma],
\eeqn
where $g_1^{\text{Full}}$ is the full computation of $g_1$, Eq.(\ref{eq:g1mix}).
The red line in the bottom plot of Fig.\ref{fig:g1} shows the relative corrections to the QCD+QED $g_1$, Eq.(\ref{eq:relativef}), using $g_1^{\text{PDF-mix}}$ for mixed-order corrections. It can be shown that the largest contribution arises from the inclusion of the mixed-order Wilson coefficient, except for values around $x\sim 0.01$, where the corrections due to PDF evolution are comparable or even dominant. 
Additionally, we compute each contribution from the Wilson coefficients separately and observe that the largest comes from the quark term $\Delta C_q^{(1,1)}\otimes(\Delta q+\Delta \bar{q})$.

\section{Conclusions}
\label{sec:conclusions}
We compute the Altarelli-Parisi splitting functions at mixed-order $\mathcal{O}(\as\al)$
using the ``Abelianization'' algorithm. Then we solve the DGLAP equations at the same accuracy, working in the Mellin 
$N\text{-space}$. 
In order to achieve this, we present two approaches to solve the matrix evolution equations for the singlet case. The first adding mixed-order corrections to a new operator
$\m{E}^{\text{MIX}}$, re-utilizing the well-known QCD and QED operators. The second is based on the Magnus expansion. Both approaches show accordance in the mixed-order corrections. The relative correction for the photon PDF reaches the percent level, even increasing to $\sim\%10$ at high values of $x$. These contributions could be significant to compute some processes sensitive to this PDF, as single-prompt photon production in electron-proton collision \cite{Rein:2024fns}.

Finally, we compute the structure function $g_1$ at mixed order, including the Wilson coefficients with $\mathcal{O}(\as\al)$ contributions. The relative corrections are found to be of the order $\mathcal{O}(10^{-4})$ for $x \simeq 0.1$, and increase to $\mathcal{O}(10^{-3})$ for $x\rightarrow1$. The largest mixed-order corrections to $g_1$ arise from the inclusion of the QCD $\otimes$ QED Wilson coefficients. Additionally, we observe that the quark channel dominates over the gluon and photon channels.
\appendix

\section{Decomposition of a generic 4x4 $\m{E}=e^{\m{R}}$}
\label{sec:MatrixR}
For an operator of the form $\m{E}=e^{\m{R}}$, where $\m{R}$ is a 4x4 matrix, we can decompose it in the projectors $\m{e}_i$ and the eigenvalues of $\m{R}$, $\lambda_i$,
\beqn
\m{E}=\m{e}_{1}e^{\lambda_1}+\m{e}_{2}e^{\lambda_2}+\m{e}_{3}e^{\lambda_3}+\m{e}_{4}e^{\lambda_4}.
\eeqn
Defining he matrix $\m{R}$ in a generic form,
\beqn
\m{R} = \left(
\begin{array}{cccc}
R_{11} & R_{12} & R_{13} & R_{14} \\
R_{21} & R_{22} & R_{23} & R_{24} \\
R_{31} & R_{32} & R_{33} & R_{34} \\
R_{41} & R_{42} & R_{43} & R_{44}
\end{array}
\right),
\label{Eq:Pqed}
\eeqn
the eigenvalues $\lambda_i$ are the roots of the polynomial,
\begin{equation}
\begin{aligned}
P_{\lambda} &= R_{14} R_{23} R_{32} R_{41} - R_{13} R_{24} R_{32} R_{41} - R_{14} R_{22} R_{33} R_{41} + R_{12} R_{24} R_{33} R_{41} \\
&\quad + R_{13} R_{22} R_{34} R_{41} - R_{12} R_{23} R_{34} R_{41} - R_{14} R_{23} R_{31} R_{42} + R_{13} R_{24} R_{31} R_{42} \\
&\quad + R_{14} R_{21} R_{33} R_{42} - R_{11} R_{24} R_{33} R_{42} - R_{13} R_{21} R_{34} R_{42} + R_{11} R_{23} R_{34} R_{42} \\
&\quad + R_{14} R_{22} R_{31} R_{43} - R_{12} R_{24} R_{31} R_{43} - R_{14} R_{21} R_{32} R_{43} + R_{11} R_{24} R_{32} R_{43} \\
&\quad + R_{12} R_{21} R_{34} R_{43} - R_{11} R_{22} R_{34} R_{43} - R_{13} R_{22} R_{31} R_{44} + R_{12} R_{23} R_{31} R_{44} \\
&\quad + R_{13} R_{21} R_{32} R_{44} - R_{11} R_{23} R_{32} R_{44} - R_{12} R_{21} R_{33} R_{44} + R_{11} R_{22} R_{33} R_{44} \\
&\quad + \left( R_{13} R_{22} R_{31} - R_{12} R_{23} R_{31} - R_{13} R_{21} R_{32} + R_{11} R_{23} R_{32} \right. \\
&\quad \left. + R_{12} R_{21} R_{33} - R_{11} R_{22} R_{33} + R_{14} R_{22} R_{41} - R_{12} R_{24} R_{41} + R_{14} R_{33} R_{41} \right. \\
&\quad \left. - R_{13} R_{34} R_{41} - R_{14} R_{21} R_{42} + R_{11} R_{24} R_{42} + R_{24} R_{33} R_{42} - R_{23} R_{34} R_{42} \right. \\
&\quad \left. - R_{14} R_{31} R_{43} - R_{24} R_{32} R_{43} + R_{11} R_{34} R_{43} + R_{22} R_{34} R_{43} + R_{12} R_{21} R_{44} \right. \\
&\quad \left. - R_{11} R_{22} R_{44} + R_{13} R_{31} R_{44} + R_{23} R_{32} R_{44} - R_{11} R_{33} R_{44} - R_{22} R_{33} R_{44} \right) \lambda \\
&\quad + \left( -R_{12} R_{21} + R_{11} R_{22} - R_{13} R_{31} - R_{23} R_{32} + R_{11} R_{33} + R_{22} R_{33} - R_{14} R_{41} \right. \\
&\quad \left. - R_{24} R_{42} - R_{34} R_{43} + R_{11} R_{44} + R_{22} R_{44} + R_{33} R_{44} \right) \lambda^2 \\
&\quad + \left( -R_{11} - R_{22} - R_{33} - R_{44} \right) \lambda^3 + \lambda^4.
\end{aligned}
\end{equation}
And the projectors $\m{e}_i$ are,
\beqn
\m{e}_1= \left( \begin{array}{cc}
-\frac{A_1 (B_2 (C_4-C_3) + B_3 (C_2-C_4) + B_4 (C_3-C_2))}{ABC} &
 -\frac{A_1 (A_2 (C_3-C_4) + A_3 (C_4-C_2) + A_4 (C_2-C_3))}{ABC}\\  
 \frac{B_1 (B_2 (C_3-C_4) + B_3 (C_4-C_2) + B_4 (C_2-C_3))}{ABC} &
 \frac{B_1 (A_2 (C_4-C_3) + A_3 (C_2-C_4) + A_4 (C_3-C_2))}{ABC} \\  
-\frac{C_1 (B_2 (C_4-C_3) + B_3 (C_2-C_4) + B_4 (C_3-C_2))}{ABC} &
 -\frac{C_1 (A_2 (C_3-C_4) + A_3 (C_4-C_2) + A_4 (C_2-C_3))}{ABC}\\  
 \frac{B_2 (C_3-C_4) + B_3 (C_4-C_2) + B_4 (C_2-C_3)}{ABC} &
 \frac{A_2 (C_4-C_3) + A_3 (C_2-C_4) + A_4 (C_3-C_2)}{ABC} \\  
\end{array}\right. \nn\\
\left.\begin{array}{cc}
 \frac{A_1 (A_2 (B_3-B_4) + A_3 (B_4-B_2) + A_4 (B_2-B_3))}{ABC} &
 \frac{A_1 (-A_2 B_3 C_4 + A_2 B_4 C_3 + A_3 B_2 C_4 - A_3 B_4 C_2 - A_4 B_2 C_3 + A_4 B_3 C_2)}{ABC}\\  
 \frac{B_1 (A_2 (B_3-B_4) + A_3 (B_4-B_2) + A_4 (B_2-B_3))}{ABC} &
 \frac{B_1 (-A_2 B_3 C_4 + A_2 B_4 C_3 + A_3 B_2 C_4 - A_3 B_4 C_2 - A_4 B_2 C_3 + A_4 B_3 C_2)}{ABC}\\  
-\frac{C_1 (A_2 (B_4-B_3) + A_3 (B_2-B_4) + A_4 (B_3-B_2))}{ABC} &
 \frac{C_1 (-A_2 B_3 C_4 + A_2 B_4 C_3 + A_3 B_2 C_4 - A_3 B_4 C_2 - A_4 B_2 C_3 + A_4 B_3 C_2)}{ABC}\\  
 \frac{A_2 (B_3-B_4) + A_3 (B_4-B_2) + A_4 (B_2-B_3)}{ABC} &
 \frac{-A_2 B_3 C_4 + A_2 B_4 C_3 + A_3 B_2 C_4 - A_3 B_4 C_2 - A_4 B_2 C_3 + A_4 B_3 C_2}{ABC}\\ 
\end{array}\right),
\eeqn
\beqn
\m{e}_2= \left( \begin{array}{cc}
 \frac{A_2 (B_1 (C_4-C_3) + B_3 (C_1-C_4) + B_4 (C_3-C_1))}{ABC} &
 \frac{A_2 (A_1 (C_3-C_4) + A_3 (C_4-C_1) + A_4 (C_1-C_3))}{ABC} \\  
-\frac{B_2 (B_1 (C_3-C_4) + B_3 (C_4-C_1) + B_4 (C_1-C_3))}{ABC} &
 -\frac{B_2 (A_1 (C_4-C_3) + A_3 (C_1-C_4) + A_4 (C3-C1))}{ABC} \\  
 \frac{C_2 (B_1 (C_4-C_3) + B_3 (C_1-C_4) + B_4 (C_3-C_1))}{ABC} &
 -\frac{C_2 (A_1 (C_4-C_3) + A_3 (C_1-C_4) + A_4 (C3-C1))}{ABC} \\  
\frac{B_1 (C_4-C_3) + B_3 (C_1-C_4) + B_4 (C_3-C_1)}{ABC} &
 \frac{A_1 (C_3-C_4) + A_3 (C_4-C_1) + A_4 (C_1-C_3)}{ABC}\\  
\end{array}\right.\nn\\
\left.\begin{array}{cc}
-\frac{A_2 (A_1 (B_3-B_4) + A_3 (B_4-B_1) + A_4 (B_1-B_3))}{ABC} &
 -\frac{A_2 (-A_1 B_3 C_4 + A_1 B_4 C_3 + A_3 B_1 C_4 - A_3 B_4 C_1 - A_4 B_1 C_3 + A_4 B_3 C_1)}{ABC}\\  
-\frac{B_2 (A_1 (B_3-B_4) + A_3 (B_4-B_1) + A_4 (B_1-B_3))}{ABC} &
 -\frac{B_2 (-A_1 B_3 C_4 + A_1 B_4 C_3 + A_3 B_1 C_4 - A_3 B_4 C_1 - A_4 B_1 C_3 + A_4 B_3 C_1)}{ABC} \\  
\frac{C_2 (A_1 (B_4-B_3) + A_3 (B_1-B_4) + A_4 (B_3-B_1))}{ABC} &
 -\frac{C_2 (-A_1 B_3 C_4 + A_1 B_4 C_3 + A_3 B_1 C_4 - A_3 B_4 C_1 - A_4 B_1 C_3 + A_4 B_3 C_1)}{ABC}\\  
 \frac{A_1 (B_4-B_3) + A_3 (B_1-B_4) + A_4 (B_3-B_1)}{ABC} &
 \frac{A_1 B_3 C_4 - A_1 B_4 C_3 - A_3 B_1 C_4 + A_3 B_4 C_1 + A_4 B_1 C_3 - A_4 B_3 C_1}{ABC}\\ 
\end{array}\right),
\eeqn

\beqn
\m{e}_3= \left( \begin{array}{cc}
-\frac{A_3 (B_1 (C_4 - C_2) + B_2 (C_1 - C_4) + B_4 (C_2 - C_1))}{ABC} &
 -\frac{A_3 (A_1 (C_2 - C_4) + A_2 (C_4 - C_1) + A_4 (C_1 - C_2))}{ABC} \\  
-\frac{B_3 (B_1 (C_4 - C_2) + B_2 (C_1 - C_4) + B_4 (C_2 - C_1))}{ABC} &
 \frac{B_3 (A_1 (C_4 - C_2) + A_2 (C_1 - C_4) + A_4 (C_2 - C_1))}{ABC} \\  
 -\frac{C_3 (B_1 (C_4 - C_2) + B_2 (C_1 - C_4) + B_4 (C_2 - C_1))}{ABC} &
 \frac{C_3 (A_1 (C_4 - C_2) + A_2 (C_1 - C_4) + A_4 (C_2 - C_1))}{ABC} \\  
\frac{B_1 (C_2 - C_4) + B_2 (C_4 - C_1) + B_4 (C_1 - C_2)}{ABC} &
 \frac{A_1 (C_4 - C_2) + A_2 (C_1 - C_4) + A_4 (C_2 - C_1)}{ABC} \\  
\end{array}\right.\nn\\
\left.\begin{array}{cc}
\frac{A_3 (A_1 (B_2 - B_4) + A_2 (B_4 - B_1) + A_4 (B_1 - B_2))}{ABC} &
 \frac{A_3 (-A_1 B_2 C_4 + A_1 B_4 C_2 + A_2 B_1 C_4 - A_2 B_4 C_1 - A_4 B_1 C_2 + A_4 B_2 C_1)}{ABC}\\  
\frac{B_3 (A_1 (B_2 - B_4) + A_2 (B_4 - B_1) + A_4 (B_1 - B_2))}{ABC} &
 \frac{B_3 (-A_1 B_2 C_4 + A_1 B_4 C_2 + A_2 B_1 C_4 - A_2 B_4 C_1 - A_4 B_1 C_2 + A_4 B_2 C_1)}{ABC} \\  
 -\frac{C_3 (A_1 (B_4 - B_2) + A_2 (B_1 - B_4) + A_4 (B_2 - B_1))}{ABC} &
 \frac{C_3 (-A_1 B_2 C_4 + A_1 B_4 C_2 + A_2 B_1 C_4 - A_2 B_4 C_1 - A_4 B_1 C_2 + A_4 B_2 C_1)}{ABC}\\  
\frac{A_1 (B_2 - B_4) + A_2 (B_4 - B_1) + A_4 (B_1 - B_2)}{ABC} &
 \frac{-A_1 B_2 C_4 + A_1 B_4 C_2 + A_2 B_1 C_4 - A_2 B_4 C_1 - A_4 B_1 C_2 + A_4 B_2 C_1}{ABC}\\ 
\end{array}\right),
\eeqn
\beqn
\m{e}_4= \left( \begin{array}{cc}
\frac{A_4 (B_1 (C_3 - C_2) + B_2 (C_1 - C_3) + B_3 (C_2 - C_1))}{ABC} &
 \frac{A_4 (A_1 (C_2 - C_3) + A_2 (C_3 - C_1) + A_3 (C_1 - C_2))}{ABC}\\  
\frac{B_4 (B_1 (C_3 - C_2) + B_2 (C_1 - C_3) + B_3 (C_2 - C_1))}{ABC} &
 -\frac{B_4 (A_1 (C_3 - C_2) + A_2 (C_1 - C_3) + A_3 (C_2 - C_1))}{ABC}\\  
-\frac{C_4 (B_1 (C_2 - C_3) + B_2 (C_3 - C_1) + B_3 (C_1 - C_2))}{ABC} &
 \frac{C_4 (A_1 (C_2 - C_3) + A_2 (C_3 - C_1) + A_3 (C_1 - C_2))}{ABC}\\  
\frac{B_1 (C_3 - C_2) + B_2 (C_1 - C_3) + B_3 (C_2 - C_1)}{ABC} &
 \frac{A_1 (C_2 - C_3) + A_2 (C_3 - C_1) + A_3 (C_1 - C_2)}{ABC}\\  
\end{array}\right.\nn\\
\left.\begin{array}{cc}
-\frac{A_4 (A_1 (B_2 - B_3) + A_2 (B_3 - B_1) + A_3 (B_1 - B_2))}{ABC} &
 -\frac{A_4 (-A_1 B_2 C_3 + A_1 B_3 C_2 + A_2 B_1 C_3 - A_2 B_3 C_1 - A_3 B_1 C_2 + A_3 B_2 C_1)}{ABC}\\  
-\frac{B_4 (A_1 (B_2 - B_3) + A_2 (B_3 - B_1) + A_3 (B_1 - B_2))}{ABC} &
 -\frac{B_4 (-A_1 B_2 C_3 + A_1 B_3 C_2 + A_2 B_1 C_3 - A_2 B_3 C_1 - A_3 B_1 C_2 + A_3 B_2 C_1)}{ABC}\\  
\frac{C_4 (A_1 (B_3 - B_2) + A_2 (B_1 - B_3) + A_3 (B_2 - B_1))}{ABC} &
 -\frac{C_4 (-A_1 B_2 C_3 + A_1 B_3 C_2 + A_2 B_1 C_3 - A_2 B_3 C_1 - A_3 B_1 C_2 + A_3 B_2 C_1)}{ABC}\\  
\frac{A_1 (B_3 - B_2) + A_2 (B_1 - B_3) + A_3 (B_2 - B_1)}{ABC} &
 \frac{A_1 B_2 C_3 - A_1 B_3 C_2 - A_2 B_1 C_3 + A_2 B_3 C_1 + A_3 B_1 C_2 - A_3 B_2 C_1}{ABC}\\ 
\end{array}\right),
\eeqn
where,
\beqn
ABC&=& A_2 B_3 C_1 - A_2 B_4 C_1 - A_1 B_3 C_2 + A_1 B_4 C_2 - A_2 B_1 C_3 + A_1 B_2 C_3 - \nn\\
  && A_1 B_4 C_3 + A_2 B_4 C_3 + A_4 (B_2 C_1 - B_3 C_1 - B_1 C_2 + B_3 C_2 + B_1 C_3 - B_2 C_3) + \\
  && A_2 B_1 C_4 - A_1 B_2 C_4 + A_1 B_3 C_4 - A_2 B_3 C_4 + \nn\\
  && A_3 (B_4 C_1 + B_1 C_2 - B_4 C_2 - B_1 C_4 + B_2 (-C_1 + C_4)),\nn
\eeqn
and,
\beqn
A_i&=&\frac{\lambda_i^3 - \lambda_i^2 (R_{22} + R_{33} + R_{44}) + \lambda_i \left(R_{22} (R_{33} + R_{44}) - R_{23} R_{32} - R_{24} R_{42} + R_{33} R_{44} - R_{34} R_{43}\right)}{\lambda_i^2 R_{41} + \lambda_i \left(R_{21} R_{42} - R_{22} R_{41} + R_{31} R_{43} - R_{33} R_{41}\right)} \nn\\
&+& \frac{- R_{22} R_{33} R_{44} + R_{22} R_{34} R_{43} + R_{23} R_{32} R_{44} - R_{23} R_{34} R_{42} - R_{24} R_{32} R_{43} + R_{24} R_{33} R_{42}}{\lambda_i^2 R_{41} + \lambda_i \left(R_{21} R_{42} - R_{22} R_{41} + R_{31} R_{43} - R_{33} R_{41}\right)} \\
&+& \frac{R_{21} R_{32} R_{43} - R_{21} R_{33} R_{42} - R_{22} R_{31} R_{43} + R_{22} R_{33} R_{41} + R_{23} R_{31} R_{42} - R_{23} R_{32} R_{41}}{\lambda_i^2 R_{41} + \lambda_i \left(R_{21} R_{42} - R_{22} R_{41} + R_{31} R_{43} - R_{33} R_{41}\right)}, \nn
\eeqn
\begin{equation}
\begin{aligned}
B_i &= \frac{\lambda_i^2 \, R_{21} + \lambda_i \left( -R_{21} (R_{33} + R_{44}) + R_{23} \, R_{31} + R_{24} \, R_{41} \right) + R_{21} \, R_{33} \, R_{44} - R_{21} \, R_{34} \, R_{43}}{\lambda_i^2 \, R_{41} + \lambda_i \left( R_{21} \, R_{42} - R_{22} \, R_{41} + R_{31} \, R_{43} - R_{33} \, R_{41} \right)} \\
&\quad - \frac{R_{23} \, R_{31} \, R_{44} - R_{23} \, R_{34} \, R_{41} + R_{24} \, R_{31} \, R_{43} - R_{24} \, R_{33} \, R_{41}}{R_{21} \, R_{32} \, R_{43} - R_{21} \, R_{33} \, R_{42}} \\
&\quad - \frac{R_{22} \, R_{31} \, R_{43} - R_{22} \, R_{33} \, R_{41} + R_{23} \, R_{31} \, R_{42} - R_{23} \, R_{32} \, R_{41}}{\lambda_i^2 \, R_{41} + \lambda_i \left( R_{21} \, R_{42} - R_{22} \, R_{41} + R_{31} \, R_{43} - R_{33} \, R_{41} \right)},
\end{aligned}
\end{equation}
\begin{equation}
\begin{aligned}
C_i &=\frac{\lambda_i^2 \, R_{31} + \lambda_i \left( R_{21} \, R_{32} - R_{22} \, R_{31} - R_{31} \, R_{44} + R_{34} \, R_{41} \right) - R_{21} \, R_{32} \, R_{44} + R_{21} \, R_{34} \, R_{42}}{\lambda_i^2 \, R_{41} + \lambda_i \left( R_{21} \, R_{42} - R_{22} \, R_{41} + R_{31} \, R_{43} - R_{33} \, R_{41} \right)} \\
&\quad + \frac{R_{22} \, R_{31} \, R_{44} - R_{22} \, R_{34} \, R_{41} - R_{24} \, R_{31} \, R_{42} + R_{24} \, R_{32} \, R_{41}}{R_{21} \, R_{32} \, R_{43} - R_{21} \, R_{33} \, R_{42}} \\
&\quad + \frac{-R_{22} \, R_{31} \, R_{43} + R_{22} \, R_{33} \, R_{41} + R_{23} \, R_{31} \, R_{42} - R_{23} \, R_{32} \, R_{41}}{\lambda_i^2 \, R_{41} + \lambda_i \left( R_{21} \, R_{42} - R_{22} \, R_{41} + R_{31} \, R_{43} - R_{33} \, R_{41} \right)}.
\end{aligned}
\end{equation}
\section{Wilson coefficients $g_1$}
\label{sec:apB}
The quark Wilson coefficient of the structure-function $g_1$ at mixed-order is given by,
\beqn
\Delta C^{(1,1)}_q&=&\frac{e_q^2 C_F}{2}\biggl[\left\{16\left(\frac{\ln {1-x}}{1-x}\right)_{+}+12\left(\frac{1}{1-x}\right)_{+}\right.\nn\\
& -&8(1+x) \ln {1-x}-4 \frac{1+x^2}{1-x} \ln {x}+2(1+x) \ln {x}-2(5+x) \nn\\
& +&\left.\left(-8 \zeta(2)+\frac{9}{2}\right) \delta(1-x)\right\} \ln{\frac{Q^2}{\mu^2_F}}\nn\\
&+&\left\{24\left(\frac{\lnn{1-x}}{1-x}\right)_{+}-12\left(\frac{\ln{1-x}}{1-x}\right)_{+}-(32 \zeta(2)\right.\\
&+&45 )\left(\frac{1}{1-x}\right)_{+}+\frac{1+x^2}{1-x}\left(4 \lnn{x}-24 \ln{x} \ln{1-x}-6 \ln{x}\right) \nn\\
& +&\frac{1+x^2}{1+x}\left(4 \lnn{x}-16 \mathrm{Li}_2(-x)-16 \ln{x} \ln{1+x}\right.\nn \\
&-&8 \zeta(2))+(1+x)\left(4 \mathrm{Li}_2(1-x)+4 \ln{x} \ln{1-x}-12 \lnn{1-x}\right.\nn\\
&-& 4 \left.\lnn{x}+16 \zeta(2)\right)+8(1+2 x) \ln{1-x}-2(1+9 x) \ln{x} \nn\\
& &\left.+2(16+11 x)+\left(40 \zeta(3)-12 \zeta(2)-\frac{51}{2}\right) \delta(1-x)\right\} \lnn{\frac{Q^2}{\mu_F^2}}\nn\\
&+&\Delta\bar{c}_{\mathrm{q}}^{(1,1) \mathrm{NS},+}-\Delta\bar{c}_{\mathrm{q}}^{(1,1) \mathrm{NS},-} \biggl],\nn
\eeqn
where $\mu_F$ is the factorization scale and, $\Delta\bar{c}_{\mathrm{q}}^{(1,1) \mathrm{NS},+}$ and $\Delta\bar{c}_{\mathrm{q}}^{(1,1) \mathrm{NS},-}$ are, 
\beqn
\Delta\bar{c}_{\mathrm{q}}^{(1,1) \mathrm{NS},+}&=&8\left(\frac{\lnnn{1-x}}{1-x}\right)_{+}-18\left(\frac{\lnn{1-x}}{1-x}\right)_{+} \nn\\
& -&(32 \zeta(2)+27)\left(\frac{\ln {1-x}}{1-x}\right)_{+}+\left(-8 \zeta(3)+36 \zeta(2)+\frac{51}{2}\right)\left(\frac{1}{1-x}\right)_{+} \nn\\
& +&\frac{1+x^2}{1-x}\bigg\{-12 S_{1,2}(1-x)+12 \operatorname{Li}_3(1-x)+48 \operatorname{Li}_3(-x)+36 \zeta(3) \nn\\
&-&6 \operatorname{Li}_2(1-x)-24 \ln {x} \operatorname{Li}_2(-x)+24 \zeta(2) \ln {x}-4 \ln {1-x} \operatorname{Li}_2(1-x) \nn\\
& +&12 \lnn{x} \ln {1-x}-14 \ln {x} \lnn{1-x}-\frac{4}{3} \lnnn{x}-\frac{3}{2} \lnn{x} \nn\\
& +&12 \ln {x} \ln {1-x} \nn\\
& +&\frac{61}{2} \ln {x}\bigg\}+(1+x)\bigg\{-4 \operatorname{Li}_3(1-x)+4 \ln {1-x} \operatorname{Li}_2(1-x) \\
& -&4 \lnnn{1-x}-4 \ln {x} \operatorname{Li}_2(1-x)-4 \zeta(2) \ln {x}+2 \ln {x} \lnn{1-x}\nn\\
& -&4 \lnn{x} \ln {1-x}+\frac{5}{3} \lnnn{x}+4 \zeta(3)\bigg\} \nn\\
& +&(1-x)\left(8 S_{1,2}(1-x)-16 \operatorname{Li}_3(-x)+8 \ln {x} \operatorname{Li}_2(-x)\right) \nn\\
& -&8\left(1+x+x^2+x^{-1}\right)\left(\operatorname{Li}_2(-x)+\ln {x} \ln{1+x}\right)-2(5+13 x) \operatorname{Li}_2(1-x) \nn\\
& -&4\left(7+2 x^2+7 x\right) \zeta(2)+2(5+7 x) \lnn{1-x}+8(1+3 x) \zeta(2) \ln {1-x} \nn\\
& +&2(5+3 x) \ln {1-x}+\left(\frac{29}{2}+4 x^2+\frac{41}{2} x\right) \lnn{x}-16(1+2 x) \ln {x} \ln {1-x} \nn\\
& +&\frac{3}{2}(3-x) \ln {x}-41-10 x+\delta(1-x)\left(6 \zeta^2(2)-78 \zeta(3)+69 \zeta(2)+\frac{331}{8}\right), \nn
\label{eq:cns+}
\eeqn
and,
\beqn 
\Delta\bar{c}_{\mathrm{q}}^{(1,1), \mathrm{NS},-}&=&\frac { 1 + x ^ { 2 } } { 1 + x } (16 \mathrm{Li}_{3}\left(\frac{1-x}{1+x}\right)-16 \mathrm{Li}_{3}\left(-\frac{1-x}{1+x}\right) \nn\\
& +&8 S_{1,2}(1-x)-16 \mathrm{Li}_{3}(1-x)-16 S_{1,2}(-x)+8 \mathrm{Li}_{3}(-x) \nn\\
& -&16 \ln {1-x} \mathrm{Li}_{2}(-x)-16 \ln{1+x} \mathrm{Li}_{2}(-x)+8 \ln {x} \mathrm{Li}_{2}(1-x) \nn\\
& +&16 \ln {x} \mathrm{Li}_{2}(-x)-16 \ln {x} \ln{1+x} \ln {1-x}+20 \lnn{2} x \ln{1+x} \nn\\
& +&4 \lnn{2} x \ln {1-x}-8 \ln {x} \lnn{2}(1+x)-8 \zeta(2) \ln {1-x} \nn\\
& -&8 \zeta(2) \ln{1+x}-2 \lnnn{3} x-8 \ln {x}+8 \zeta(3) ) \\
& +&(1+x)(16 S_{1,2}(-x)-8 \operatorname{Li}_{3}(-x)+16 \ln{1+x} \mathrm{Li}_{2}(-x). \nn\\
& +&8 \zeta(2) \ln{1+x}+8 \ln {x} \lnn{2}(1+x)-4 \lnn{2} x \ln{1+x}+8 \mathrm{Li}_{2}(1-x) \nn\\
& +&8 \ln {x} \ln {1-x}-8 \zeta(3))+(1-x)(16 \ln{1-x}+30) \nn\\
&+&(x^{2}+x^{-1})(\operatorname{Li}_{2}(-x)+\ln {x} \ln{1+x})-4(2+x^{2}+x) \lnn{2} x \nn\\
& +&4(1+2 x^{2}-x) \zeta(2)+(6+38 x) \ln {x} . \nn
\label{eq:cns-}
\eeqn
where $\operatorname{Li}_{n}(x)$ and $S_{n,p}(x)$ denote characteristic polylogarithmic functions that arise in higher-order radiative corrections.
The gluon Wilson coefficient is,
\beqn
\Delta C^{(1,1)}_g&=&\frac{e_q^2 T_R}{4}   \Bigg\{\{8(1-2 x)\left(-\operatorname{Li}_2(1-x)-2 \lnn{1-x}+3 \ln {x} \ln{1-x}-\lnn{x}+4 \zeta(2)\right) \nn \\
& +&4(17-20 x) \ln {1-x}-4(12-8 x) \ln{x}-68+52 x\} \ln{\frac{Q^2}{\mu^2_F}} \nn\\ 
&+& \{4(1-2x)(-2\ln{1-x}+\ln{x})+6\}\lnn{\frac{Q^2}{\mu^2_F}}\\
&+&(1-2 x)\bigg(32 \operatorname{Li}_3(1-x)-16 \ln {1-x} \operatorname{Li}_2(1-x)-8 \ln {x} \operatorname{Li}_2(1-x) \nn\\
& -&24 \zeta(2) \ln {x}-\frac{20}{3} \lnnn{1-x}+16 \ln {x} \lnn{1-x}-16 \lnn{x} \ln {1-x}\nn\\
&+&\frac{10}{3} \lnnn{x}\bigg)-16\left(1+x^2+2 x\right)\left(4 S_{1,2}(-x)+4 \ln{1+x} \operatorname{Li}_2(-x)\right. \nn\\
&&\left.+2 \ln {x} \lnn{1+x}-\lnn{x} \ln{1+x}+2 \zeta(2) \ln{1+x}\right)-32\left(1+x^2\right. \nn\\
&-&6 x) \operatorname{Li}_3(-x)+8\left(1+4 x^2-2 x\right) S_{1,2}(1-x) \nn\\
& +&\frac{16}{3}\left(13 x^2+12 x+4 x^{-1}\right)\left(\operatorname{Li}_2(-x)+\ln {x} \ln{1+x}\right) \nn\\
&+ & 4(5-12 x) \mathrm{Li}_2(1-x)+32\left(1+x^2-2 x\right) \ln {x} \mathrm{Li}_2(-x)+\frac{1}{3}\left(123-104 x^2-48 x\right) \lnn{x} \nn\\
& -&(88-96 x) \ln {x} \ln {1-x}+6(9-12 x) \lnn{1-x}-32 \zeta(2) x^2 \ln {1-x} \nn\\
& -&4\left(31-4 x^2-26 x\right) \ln {1-x}+\frac{1}{3}\left(416-48 x^2-274 x\right) \ln{x}-8\left(5-4 x^2-26 x\right) \zeta(3) \nn\\
& -&  \frac{4}{3}\left(81-52 x^2-108 x\right) \zeta(2)+\frac{2}{3}(233-239 x) \nn\\
&+&16(1+2 z)(2 \operatorname{Li}_2(1-z)+2 \ln {z} \ln {1-z}-\lnn{z})\nn\\
&+&96(1-z) \ln{1-z}-(144+64 z) \ln{z}-304(1-z) 
\Bigg\},\nn
\eeqn
and finally, the photon Wilson coefficient is,
\beqn
\Delta C_\gamma^{(1,1)}=  \frac{C_A C_F}{T_R} \Delta C_g^{(1,1)}.
\eeqn
\begin{acknowledgements}
This work is partially supported by CONICET. We thank Stefano Forte and Felix Hekhorn for for many valuable and insightful discussions.
\end{acknowledgements}
\bibliography{biblio.bib}   

\end{document}